\documentclass[14pt]{article}

\usepackage{PRIMEarxiv}
\usepackage[utf8]{inputenc} % allow utf-8 input
\usepackage[T1]{fontenc}    % use 8-bit T1 fonts
\usepackage{hyperref}       % hyperlinks
\usepackage{url}            % simple URL typesetting
\usepackage{booktabs}       % professional-quality tables
\usepackage{amsfonts}       % blackboard math symbols
\usepackage{amsmath}
\usepackage{bm}
\usepackage{easyReview}
\usepackage{bbm}
\usepackage{nicefrac}       % compact symbols for 1/2, etc.
\usepackage{microtype}      % microtypography
\usepackage{lipsum}
\usepackage{fancyhdr}       % header
\usepackage{graphicx}       % graphics
\usepackage{xcolor}
\graphicspath{{media/}}     % organize your images and other figures under media/ folder
\usepackage{natbib}
\usepackage{float}
\usepackage{mathtools}
\usepackage{makeidx}
\makeindex
\usepackage[thinc]{esdiff}
\usepackage{csquotes}
\usepackage{easyReview}
\usetikzlibrary{positioning,shapes,calc,backgrounds}
\definecolor{amaranth}{rgb}{0.9, 0.17, 0.31}

\definecolor{hanblue}{rgb}{0.27, 0.42, 0.81}

\definecolor{indianyellow}{rgb}{0.89, 0.66, 0.34}

\usepackage{enumitem}

\definecolor{blue_user}{HTML}{48ADFF}
\definecolor{yellow_user}{HTML}{FFDC8E}
\definecolor{blueuser}{HTML}{48ADFF}
\definecolor{yellowuser}{HTML}{FFDC8E}

\usetikzlibrary{shapes.geometric, arrows, positioning}
\usepackage{rotating}
\usepackage{array}

\tikzstyle{mainblock} = [
    rectangle, draw, text width=10cm, align=center,
    minimum height=4em, fill=blueuser!30
]

\tikzstyle{subblock} = [
    rectangle, draw, text width=12cm, align=left,
    minimum height=3em, inner sep=6pt
]

\tikzstyle{startend} = [
    ellipse, draw, text width=1cm, align=center,
    minimum height=3em, fill=yellowuser!50
]

\tikzstyle{arrow} = [dashed,->,thick]

\usepackage{adjustbox}
\title{Beyond Linearity and Time-Homogeneity: \\
Relational Hyper Event Models \\with Time-Varying Non-Linear Effects}

\author{Martina Boschi\thanks{Corresponding author} \\
\texttt{\url{martina.boschi@usi.ch}} \\
Università della Svizzera italiana \\
Via la Santa 1, 6962 \\
Lugano-Viganello, Switzerland
\And 
J\"urgen Lerner \\
\texttt{\url{juergen.lerner@uni-konstanz.de}} \\
University of Konstanz \\
Konstanz, Germany
\And
Ernst C. Wit \\
\texttt{\url{ernst.jan.camiel.wit@usi.ch}} \\
Università della Svizzera italiana \\
Lugano, Switzerland}

\begin{document}

    % TITLE PAGE
    
\maketitle%
\begingroup
\renewcommand{\thefootnote}{} % make footnote unnumbered
\footnote{\tiny \textbf{List of Abbreviations}: \begin{itemize}[topsep=0pt, partopsep=0pt, itemsep=0.5pt, left=0pt]
    \item REM: Relational Event Model
    \item RHEM: Relational Hyper Event Model
    \item DyNAM: Dynamic Network Actor Model
    \item TVE: Time-Varying Effect
    \item NLE: Non-Linear Effect
    \item ERGM: Exponential Random
Graph Model
    \item LE: Linear Effect
    \item TVNLE: \textit{Jointly} Time-Varying Non-Linear Effect
    \item AAS: Average Author Score
    \item RPI: Referenced Papers’ Impact
    \item SCAM: Shape Constrained Additive Models
    \item TPRS: Thin Plate Regression Spline
    \item TPS: Thin Plate Spline    
    \item GLM: Generalized Linear Model
    \item GAM: Generalized Additive Model 
    \item ECDF: Empirical Cumulative Distribution Function
    \item KS: Kolmogorov–Smirnov
    \item AIC: Akaike Information Criterion
    \item LogLik: Log-Likelihood
\end{itemize}}
\endgroup

\begin{abstract}

   Recent technological advances have made it easier to collect large and complex networks of time-stamped relational events connecting two or more entities. Relational hyper-event models (RHEMs) aim to explain the dynamics of these events by modeling the event rate as a function of statistics based on past history and external information.
  
   However, despite the complexity of the data, most current RHEM approaches still rely on a linearity assumption to model this relationship. In this work, we address this limitation by introducing a more flexible model that allows the effects of statistics to vary non-linearly and over time. While time-varying and non-linear effects have been used in relational event modeling, we take this further by modeling joint time-varying and non-linear effects using tensor product smooths.
  
   We validate our methodology on both synthetic and empirical data. In particular, we use RHEMs to study how patterns of scientific collaboration and impact evolve over time. Our approach provides deeper insights into the dynamic factors driving relational hyper-events, allowing us to evaluate potential non-monotonic patterns that cannot be identified using linear models.

\end{abstract} 

\keywords{relational hyper event model (RHEM); generalized additive model (GAM); tensor product smooths; time-varying effect; non-linear effect; non-monotonic patterns;}

\newpage
    
    % INTRODUCTION
    %----------------------------------------------------------------------------%
% INTRODUCTION
\section{Introduction}
\label{sec_introduction}
%----------------------------------------------------------------------------%

%----------------------------------------------------------------------------%
% Introducing the context: from some examples to the general concept
Activities such as sending e-mails \citep{perry2013point, boschi2024goodness}, executing financial transactions \citep{bianchi2023ties}, attending meetings \citep{lerner2021dynamic}, citing academic references \citep{filippi2024stochastic, lerner2024relational}, co-participating in criminal activities \citep{bright2024examining}, and collaborating in cultural production \citep{burgdorf2024communities} represent distinct social phenomena, that share a fundamental characteristic: they can all be represented as \textit{temporal interactions}\index{temporal interaction} involving two or more participants. This relational perspective lies at the heart of social network analysis and distinguishes it from other analytical frameworks \citep{wasserman1994social}. In an event-based representation of temporal networks, social dynamics are modeled as collections of time-stamped edges that capture the evolving structure of relationships over time \citep{lambiotte2016guide}. The aforementioned examples often involve relationships that extend beyond simple dyadic interactions, consisting of complex and multi-actor activities occurring at specific points in time. We refer to these richer forms of interaction as \textit{hyperevents}\index{hyperevent} \citep{lerner2021dynamic} -- mathematically represented as time-stamped \textit{hyperedges}\index{hyperedge} \citep{berge1989hypergraphs}, i.e., sets of nodes of arbitrary size. Modeling these hyperevents involves accounting for the polyadic nature of such interactions \citep{seidman1981structures}.

%----------------------------------------------------------------------------%
% State of the art
An increasing number of studies have focused on describing temporal hypergraphs using temporal motifs and relative counts \citep{paranjape2017motifs, wang2020efficient, lee2023temporal}. However, such approaches do not model the dynamics of how hyperedges emerge. Some progress has been made in detecting changes in the dynamics of temporal hypergraphs driven by specific factors \citep{Zou2017Modeling}. Notably, \citet{benson2018sequences} propose a generative stochastic model that predicts and explains the occurrence of sets of entities -- interpretable as hyperevents -- based on previously observed subsets. This growing interest in modeling complex, time-stamped group interactions has led to the extension of traditional \textit{Relational Event Models}\index{relational event model} (REMs), which primarily focus on sequences of events involving a sender and a receiver \citep{butts2008relational, bianchi2024relational}, into the hypergraph domain. \textit{Relational Hyper Event Models}\index{relational hyper event model} (RHEMs) aim to capture the complex dynamics of polyadic events, conceived as interactions involving multiple participants simultaneously \citep{lerner2021dynamic, lerner2023relational}. Going beyond the initial proposal by \citet{perry2013point} for modeling \textit{multi-cast interactions}\index{multi-cast interactions}, RHEMs provide a comprehensive framework for understanding the dynamics of social-relational hyper-event processes. {This modeling approach allows us to address several important empirical questions. Consider co-offending, which can be treated as a relational hyper-event because it involves multiple individuals. Relational hyper-event models can explain co-offending dynamics as a function of the age and sex of the individuals involved, homophily with respect to these characteristics, and prior solo offending activity \citep{bright2024examining}. The same idea extends naturally to other social phenomena discussed at the beginning of this introduction. For example: How do reciprocity and transitive closure influence email exchanges? How does transaction amount or country of the beneficiary account influence the rate of bank transfers? Such questions can be addressed by modeling the intensity of the corresponding social processes as a function of observed covariates.}

In relational hyper event modeling, intensity processes are indeed modeled as functions of higher-order covariates \citep{lerner2023relational}. Recent RHEM formulations allow for the analysis of dependencies among nodes across different network modes, accommodating entities with inherently distinct characteristics -- such as authors (scientists) and references (scientific papers) within a publication network \citep{lerner2024relational}. The computational efficiency in calculating hyperedge covariates, enabled by the open-source software \texttt{eventnet}\index{eventnet} \citep{lerner2020reliability, lerner2023relational}, makes RHEMs a practical tool in social network analysis. This facilitates the investigation of dynamic network effects -- such as preferential attachment, (partial) repetition, and triadic closure -- that may hold theoretical or empirical significance.

%----------------------------------------------------------------------------%
% Limitations of the state of the art
Despite significant advancements in the RHEM literature, current formulations remain constrained by a fundamental limitation: they assume \textit{linearity} in hyperevent effects. This reliance introduces two key drawbacks. First, it overlooks the temporal variability of effects. In temporal networks -- where endogenous and complex dynamics are inherently time-dependent -- several REM formulations have underscored the importance of incorporating time-varying effects into empirical applications \citep{juozaitiene2023analysing, boschi2023smooth}. 
{The work of \citet{kamalabad2023point} further motivates the need to account for changes in the impact of covariates in REMs. This need arises due to potential changes in the social interaction behavior of the entities in a temporal network. They test specific time points at which such changes in the parameters in the model formulation may be necessary. Within the broader framework of modeling relational event data, time-varying effects have also been incorporated into Dynamic Network Actor Models (DyNAMs) \citep{uzaheta2024modeling}. The underlying motivation is that the temporal network may occupy multiple states over time, and that the influence of network features on actor choices may vary across these states.}

Second, the linearity assumption restricts the flexibility of existing REM frameworks, limiting their capacity to capture how covariate effects vary as a function of their values \citep{bauer2022smooth, filippi2024modeling, filippi2024analyzing}. This limitation is particularly relevant when covariates are evaluated in terms of their \textit{internal time}\index{internal time} \citep{juozaitiene2024nodal, amati2024goodness}, a common scenario for many endogenous covariates in relational event modeling. Mechanisms such as \textit{reciprocity}\index{reciprocity} and \textit{repetition}\index{repetition} -- computed from previous events in the opposite and same direction, respectively -- can be more effectively assessed by considering the time intervals between current and past relevant events. While these prior interactions influence the rate of events, their impact typically weakens with time -- a phenomenon referred to as \textit{forgetfulness} \citep{juozaitiene2024time}. To capture this decaying influence, it is advantageous to model these covariates, measured as internal times, using non-linear effects. This allows the decay patterns to be learned directly from the data, rather than imposing a rigid, prespecified functional form. Such flexibility is helpful because relying on an incorrect decay function can lead to erroneous conclusions \citep{arena2023fast}. A related and equally important phenomenon is \textit{saturation}\index{saturation}, which arises when individuals face cognitive overload and become unable to effectively process additional information \citep{atienza2025modeling}. Within the RHEM framework, saturation can be modeled by allowing the influence of certain drivers on the event rate to exhibit a non-linear pattern -- initially increasing with exposure but eventually reaching a plateau or declining.
{Indeed, we further emphasize that, for a linear model, the issue is not only non-linearity but especially non-monotonicity. A linear function can oversimplify a monotonic non-linear trend while still capturing its overall direction. However, it is inherently unable to represent a non-monotonic pattern. In some cases, for example, over-exposure may even reverse the direction, making it counter-effective. A linear model would be unable to reflect such a reversal and would instead provide a single estimate that ``averages'' the increasing and decreasing tendencies observed in the data.}

%----------------------------------------------------------------------------%
% Proposal
This paper addresses the primary limitation of current RHEM formulations by incorporating recent advancements from the REM literature, particularly regarding \textit{time-varying effects}\index{effect!time-varying effect} (TVEs) and \textit{non-linear effects}\index{effect!non-linear effect} (NLEs). However, our contribution goes beyond a straightforward adaptation of TVEs and NLEs to the context of relational hyper event modeling. We introduce a novel framework that enables effects to be \textit{jointly time-varying and non-linear}\index{effect!joint time-varying non-linear effect}. This is accomplished through the use of \textit{tensor product bases}\index{smooth function!tensor product bases}, which allow for the construction of smooth functions involving multiple covariates simultaneously \citep{wood2017generalized}. As a result, the proposed method supports the representation of a smooth interaction term, defined on the Cartesian product of time and the covariate of interest.

%----------------------------------------------------------------------------%
% Outline of the paper
{The rest of the paper is organized as follows: We expand the motivation for this contribution in Section \ref{sec_motivation}.} The methodological section of the paper is presented in Section~\ref{sec_dynamic_hypernetwork_modeling}, which outlines the modeling framework and inference techniques for relational hyperevent sequences.
Section~\ref{sec_simulation_study} describes a simulation study using synthetic relational hyperevent data. The results show that when the true data-generating process involves time-varying effects, non-linear effects, or the two of them jointly, the linear model fails to capture these dynamics. In contrast, the proposed model successfully recovers these effects. Moreover, when the assumption of linearity holds, the proposed framework naturally reduces to the correct linear specification, recovering it as a special case {and without overfitting}. Section~\ref{sec_empirical_application} presents empirical results on scientific collaboration, extending the models developed by \citet{lerner2024relational}. In this application\footnote{The code required to implement the empirical analysis is provided in the Supplementary Material.}, we find evidence for effects that are linear and fixed in time, but also -- and especially -- non-linear, and jointly time-varying non-linear, demonstrating the relevance of this approach in empirical hyperevent applications that take place over a long period -- 80 years in our case. {For instance, we identify non-monotonic effects -- such as for author self-citation -- whose influence on publication rates changes over time and in some time-slots end up being clearly non-monotonic. Such patterns cannot be captured by linear specifications and motivate extending relational hyper event models beyond linear formulations.}

    % MOTIVATION
    \section{Motivation}
\label{sec_motivation}

{Science and technology are key components in shaping the direction of innovation \citep{wang2021probing}. Innovation is a complex phenomenon, and defining proper measures of innovation has long been a central interest among researchers \citep{todorov2024economies}. Among the earliest measures of \textit{technological} innovation has been the simple count of patents \citep{griliches1998patent}, although this alone fails to capture the interconnectedness of innovation \citep{jaffe2002patents}. More recent approaches focus on patent citations, based on the idea that if a patent has been published, it likely builds on prior patented work referenced in the cited patents \citep{filippi2024stochastic}, thereby identifying a ``knowledge flow.'' %Recognizing the importance of networks in technological innovation has indeed become almost self-evident \citep{breschi2004knowledge}. 
Similarly, \textit{scientific} innovation has been described as a process where citations tracks the diffusion of innovation \citep{zhai2018measuring}. Although citation networks have been described as static network, e.g., using Exponential Random Graph Models (ERGMSs) \citep{chakraborty2020patent}, the dynamic flow  of citations is better captured using dynamic network modeling techniques such as REMs \citep{artico2023dynamic}.}

{Given the complex nature of the citation process, the need of moving beyond a linear REM was recognized early on \citep{bauer2022smooth, filippi2024analyzing}. This allows the discovery of interesting non-monotonic patterns that linear models would fail to capture. For example, \citet{filippi2024stochastic}, based on an analysis of United States Patent and Trademark Office patent citation data, detect that the citation rate of an individual patent shows a clear peak appearing shortly after the year 2000, in contrast to previously reported steady increases in the patent citation process. Additionally, the temporal lag at which future patents are most likely to be cited also exhibits a peak, occurring around five years.}

{\citet{golosovsky2019citation} proposes a Hawkes self-exciting, non-linear process to model the latent citation rate of a paper over time.  They demonstrate the non-linearity of citation dynamics and justify it as closely related to network topology. In particular, the effect of a paper’s citation history on the rate of new citations must itself depend non-linearly on its past, reflecting the very different trajectories of moderately versus highly cited papers. \citet{singh2023forgotten} consider another phenomenon called  \textit{citational amnesia}, defined as the tendency to ``forget'' to cite high-quality older work. Citational amnesia affects future citations in two distinct ways. First, the recency of past publications affects how likely they are to be cited. Second, acknowledging that the impact of historical publication activity may itself change as the scientific ecosystem evolves is essential. Allowing for both non-linearity (to capture non-linear effects of past activity) and time-variation (to incorporate the historical drivers of citation amnesia) is therefore crucial.}

{Non-linear effects in relational dynamics happens in various situations. \citep{juozaitiene2024time} show the importance of non-linear reciprocity and transitive closure effects in various communication networks. \citep{boschi2024goodness} examine email communication data in a manufacturing company, demonstrating that the tendency to reply to an email depend in a strongly non-linear way on the time since the last email received from that individual. 
An example on time-varying effects in relational event models can be found in the effect of trade on alien species diffusion process over the past 150 years \citep{juozaitiene2023analysing,boschi2023smooth}. The effect of trade in this context is shown to vary over time: the diffusive effect of trade has gradually been decreasing over time. A possible explanation is that the nature of traded goods has changed, becoming less closely relevant for the invasion processes in recent years. The work of \citet{kamalabad2023point} further underlines the need for testing the presence of change points and the consequent modification of the model formulation to account for time-varying effect. \citet{uzaheta2024modeling} extend a special class of dynamic network models, DyNAMs, to allow for time-varying parameters, thereby accounting for  changes in the system over time. Specifically, they assume that the state of the network may change over time, and depending on the state, the network features may affect the rate and choice sub-models that constitute a DyNAM differently. They apply their method to evaluate communication patterns in a MIT Media Lab’s ``Social Evolution'' study.}
    
    % DYNAMIC HYPERNETWORK MODELING
    %----------------------------------------------------------------------------%
% DYNAMIC HYPERNETWORK MODELING
\section{Dynamic Hypernetwork Modeling}
\label{sec_dynamic_hypernetwork_modeling}
%----------------------------------------------------------------------------%

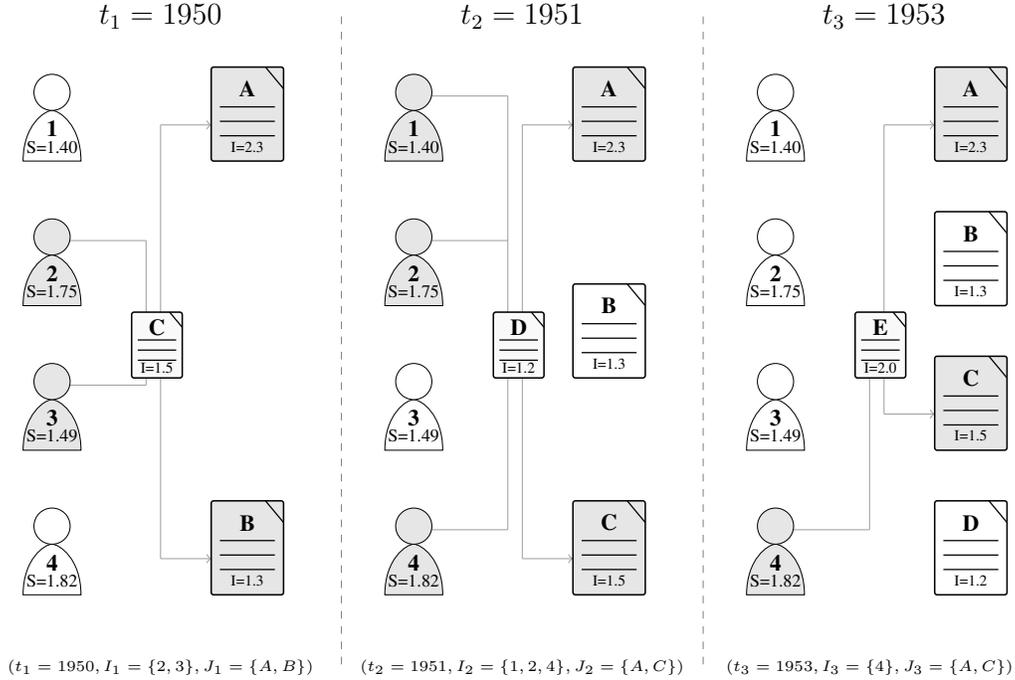
\begin{figure}[tb]
\centering
% ---------- reusable pics definitions ----------
\tikzset{
  pics/user/.style args={#1/#2/#3}{
    code={
      % Body
      \draw[fill=#2, draw=black] 
        (-0.4,0) .. controls (-0.4,0.5) and (-0.1,0.7) .. (0,0.7)
        .. controls (0.1,0.7) and (0.4,0.5) .. (0.4,0) -- cycle;
      % Text inside body
      \node[font=\small] at (0,0.45) {\textbf{#1}};
      \node[font=\scriptsize] at (0,0.2) {S=#3};
      % Head
      \filldraw[fill=#2, draw=black] (0,0.95) circle (0.25cm);
    }
  },
  pics/paper/.style args={#1/#2/#3}{
    code={
      \fill[#2] (0,0) rectangle (1.0,1.3);
      \draw[line width=0.6pt, rounded corners=1pt] (0,0) rectangle (1.0,1.3);

      \node[font=\small] at (0.5,1) {\textbf{#1}};
      \node[font=\tiny] at (0.5,0.2) {I=#3};
      % Lines
      \foreach \y in {0.75,0.55,0.35} {
        \draw[line width=0.5pt] (0.12,\y) -- (0.88,\y);
      }
      % Folded corner
      \draw[line width=0.5pt] (0.75,1.3) -- (1.0,1.0);
    }
  }
}
% ---------- end pics ----------

\resizebox{!}{0.4\textheight}{
\begin{tikzpicture}[user/.style={scale=1.0}, paper/.style={scale=1.0}]

% Users y-coordinates
\coordinate (U1) at (0,0);
\coordinate (U2) at (0,-2);
\coordinate (U3) at (0,-4);
\coordinate (U4) at (0,-6);

% Horizontal offsets for t=1,2,3
\def\xA{0}   % t=1
\def\xB{5}   % t=2
\def\xC{10}  % t=3

% ===== t = 1 =====
\draw[-, gray!75] ($(U2)+(\xA,0)+(0,0.9)$) --  ($(U2)+(\xA+1.3,0)+(0,0.9)$);
\draw[-, gray!75] ($(U2)+(\xA+1.3,0)+(0,0.9)$) --  ($(\xA+1.3,-3.5)+(0,0.9)$);

\draw[-, gray!75] ($(U3)+(\xA,0)+(0,0.9)$) --  ($(U3)+(\xA+1.3,0)+(0,0.9)$);
\draw[-, gray!75] ($(U3)+(\xA+1.3,0)+(0,0.9)$) --  ($(\xA+1.3,-3.5)+(0,0.9)$);

\draw[-, gray!75] ($(\xA+1.5,-3.5)+(0,0.9)$) -- ($(\xA+1.5,0.5)$);
\draw[->, gray!75] ($(\xA+1.5,0.5)$) -- ($(\xA+2.2,0.5)$);

\draw[-, gray!75] ($(\xA+1.5,-3.5)+(0,0.9)$) -- ($(\xA+1.5,-5.5)$);
\draw[->, gray!75] ($(\xA+1.5,-5.5)$) -- ($(\xA+2.2,-5.5)$);

% ===== t = 2 =====
\draw[-, gray!75] ($(U1)+(\xB,0)+(0,0.9)$) --  ($(U1)+(\xB+1.3,0)+(0,0.9)$);
\draw[-, gray!75] ($(U1)+(\xB+1.3,0)+(0,0.9)$) -- ($(\xB+1.3,-3.5)+(0,0.9)$);
\draw[-, gray!75] ($(U2)+(\xB,0)+(0,0.9)$) -- ($(U2)+(\xB+1.3,0)+(0,0.9)$);
\draw[-, gray!75] ($(U4)+(\xB,0)+(0,0.9)$) -- ($(U4)+(\xB+1.3,0)+(0,0.9)$);
\draw[-, gray!75] ($(U4)+(\xB+1.3,0)+(0,0.9)$) -- ($(\xB+1.3,-3.5)+(0,0.9)$);

% Blue paper D → yellow papers A and C
\draw[-, gray!75] ($(\xB+1.5,-3.5)+(0,0.9)$) -- ($(\xB+1.5,0.5)$);
\draw[->, gray!75] ($(\xB+1.5,0.5)$) -- ($(\xB+2.2,0.5)$);

\draw[-, gray!75] ($(\xB+1.5,-3.5)+(0,0.9)$) -- ($(\xB+1.5,-5.5)$);
\draw[->, gray!75] ($(\xB+1.5,-5.5)$) -- ($(\xB+2.2,-5.5)$);

% ===== t = 3 =====
\draw[-, gray!75] ($(U4)+(\xC,0)+(0,0.9)$) -- ($(U4)+(\xC+1.3,0)+(0,0.9)$);
\draw[-, gray!75] ($(U4)+(\xC+1.3,0)+(0,0.9)$) -- ($(\xC+1.3,-3.5)+(0,0.9)$);

\draw[-, gray!75] ($(\xC+1.5,-3.5)+(0,0.9)$) -- ($(\xC+1.5,0.5)$);
\draw[->, gray!75] ($(\xC+1.5,0.5)$) -- ($(\xC+2.2,0.5)$);

\draw[-, gray!75] ($(\xC+1.5,-3.5)+(0,0.9)$) -- ($(\xC+1.5,-3.5)$);
\draw[->, gray!75] ($(\xC+1.5,-3.5)$) -- ($(\xC+2.2,-3.5)$);

% ---------------- t=1 ----------------
\pic at ($ (U1)+(\xA,0) $) {user={1/white/1.40}};
\pic at ($ (U2)+(\xA,0) $) {user={2/gray!20/1.75}};
\pic at ($ (U3)+(\xA,0) $) {user={3/gray!20/1.49}};
\pic at ($ (U4)+(\xA,0) $) {user={4/white/1.82}};

% Papers t=1
\pic at ($ (\xA+2.2,0) $) {paper={A/gray!20/2.3}};
\pic at ($ (\xA+2.2,-6) $) {paper={B/gray!20/1.3}};
\pic at ($ (\xA+1.1,-3) $) [scale=0.7] {paper={C/gray!5/1.5}};

% ---------------- t=2 ----------------
\pic at ($ (U1)+(\xB,0) $) {user={1/gray!20/1.40}};
\pic at ($ (U2)+(\xB,0) $) {user={2/gray!20/1.75}};
\pic at ($ (U3)+(\xB,0) $) {user={3/white/1.49}};
\pic at ($ (U4)+(\xB,0) $) {user={4/gray!20/1.82}};

% Papers t=2
\pic at ($ (\xB+2.2,0) $) {paper={A/gray!20/2.3}};
\pic at ($ (\xB+2.2,-3) $) {paper={B/white/1.3}};
\pic at ($ (\xB+2.2,-6) $) {paper={C/gray!20/1.5}};
\pic at ($ (\xB+1.1,-3) $) [scale=0.7] {paper={D/gray!5/1.2}};

% ---------------- t=3 ----------------
\pic at ($ (U1)+(\xC,0) $) {user={1/white/1.40}};
\pic at ($ (U2)+(\xC,0) $) {user={2/white/1.75}};
\pic at ($ (U3)+(\xC,0) $) {user={3/white/1.49}};
\pic at ($ (U4)+(\xC,0) $) {user={4/gray!20/1.82}};

% Papers t=3
\pic at ($ (\xC+2.2,0) $) {paper={A/gray!20/2.3}};
\pic at ($ (\xC+2.2,-2) $) {paper={B/white/1.3}};
\pic at ($ (\xC+2.2,-4) $) {paper={C/gray!20/1.5}};
\pic at ($ (\xC+2.2,-6) $) {paper={D/white/1.2}};
\pic at ($ (\xC+1.1,-3) $) [scale=0.7] {paper={E/gray!5/2.0}};

% ---------------- Vertical lines ----------------
\draw[dashed, gray] (4,2) -- (4,-7);
\draw[dashed, gray] (9,2) -- (9,-7);

% Label for t=1
\node[font=\large] at (\xA+1.5,2) {$t_1=1950$};
\node[font=\tiny] at (\xA+1.5,-7) {$(t_1=1950, I_1=\{2,3\}, J_1=\{A,B\})$};

% Label for t=2
\node[font=\large] at (\xB+1.5,2) {$t_2=1951$};
\node[font=\tiny] at (\xB+1.5,-7) {$(t_2=1951, I_2=\{1,2,4\}, J_2=\{A,C\})$};

% Label for t=3
\node[font=\large] at (\xC+1.5,2) {$t_3=1953$};
\node[font=\tiny] at (\xC+1.3,-7) {$(t_3=1953, I_3=\{4\}, J_3=\{A,C\})$};

\end{tikzpicture}
}
{
\caption{\textbf{Motivating example: publication events.} This picture represent the first three events of a synthetic relational hyperevent sequence of paper publications. Each directed hyperevent consists of a paper, written by a team of authors (senders) and citing a set of references (receivers). Each hyperevent comes with a time-stamp, reporting the publication date. Only when the paper is published, it can be cited. This is why published paper appears as potential paper in the following time stamp. Furthermore, each actor is associated with an score (S), while each paper has a numeric impact associated (I).}
}
\label{fig:example_introduction}
\end{figure}

% Hypernetworks, hyperevent, hyperedges
A \textit{dynamic hypernetwork}\index{dynamic hypernetwork} is a temporal system in which hyperevents \enquote{occur}. These systems are also called \textit{relational hypernetworks}\index{relational hypernetwork}. They are made up of a sequence of hyperevents, written as:
\begin{equation}\label{eq_hyperevent_sequence}  
    E = \{(t_m, I_m, J_m)\}, \quad m = 1, \dots, n,
\end{equation}  
where \(t_m\) is the time when a group of \textit{senders}\index{sender} \(I_m\) interacts with a group of \textit{receivers}\index{receiver} \(J_m\). All participants are nodes in a vertex set \(V\) of a \textit{temporal hypergraph}\index{temporal hypergraph}. Depending on the context, senders and receivers may belong to the same or different groups. If they belong to the same group, the system is a \textit{one-mode system}\index{relational hypernetwork!one-mode system}. If they play different roles, the system is a \textit{two-mode system}\index{relational hypernetwork!two-mode system}. In the two-mode case, the vertex set \(V\) of the \textit{two-model temporal hypergraph}\index{temporal hypergraph!two-model temporal hypergraph} is divided into two disjoint sets: \(V^I\) for possible senders and \(V^J\) for possible receivers. If entities join or leave the system over time, these sets can vary with time and are written as \(V^I_t\) and \(V^J_t\). In our empirical study, we focus on a two-mode system related to scientific publications, where a group of authors (senders) cite a group of previously published papers (receivers). 

{Throughout this paper, we will refer to a simple synthetic example to introduce our methodological contributions from a practical point of view, relying on the scientific empirical application presented later. Figure~\ref{fig:example_introduction} reports the first three publication events in a synthetic event sequence \(E\) spanning from 1950 to 2021. At each time point, we observe a set \(V^I\) of potential senders and a set \(V^J\) of potential receivers. Specifically, we consider a set of hypothetical hyperevents as shown also in Table~\ref{tab:hyperevents}. This example is useful for highlighting the importance of defining a proper time-evolving risk set: indeed, a paper cannot cite another paper that has not yet been published.}

\begin{table}[tb]
\centering
\adjustbox{width=\textwidth}{
\begin{tabular}{l l l l l l l}
\hline
Index $m$ & Time $t_m$ & Potential Authors $V^I_{t_m}$ & Observed Authors $I_m$ & Article $j_m$ & Potential Citations $V^J_{t_m}$ & Observed Citations $J_m$ \\
\hline
1 & 1950 & $\{1,2,3,4\}$ & $\{2,3\}$ & $C$ & $\{A,B\}$ & $\{A,B\}$ \\
2 & 1951 & $\{1,2,3,4\}$ & $\{1,2,4\}$ & $D$ & $\{A,B,C\}$ & $\{A,C\}$ \\
3 & 1953 & $\{1,2,3,4\}$ & $\{4\}$ & $E$ & $\{A,B,C,D\}$ & $\{A,C\}$ \\
\vdots & \vdots & \vdots & \vdots & \vdots & \vdots & \vdots \\
\hline
\end{tabular}
}
\caption{Collection of hyperevents $\{ (t_m, I_m, J_m)~|~m=1,\ldots,2000\}$ representing published papers whereby a set of co-authors $I_m$ cites a set of papers $J_m$ in a published article $j_m$. \label{tab:hyperevents}}
\end{table}

The notation \((t_m, I_m, J_m)\) in Equation~\ref{eq_hyperevent_sequence} naturally describes \textit{directed hyperevents}\index{hyperevent!directed hyperevent}, where sender and receiver groups are distinct. {The three directed hyperevents described in Table~\ref{tab:hyperevents} can be represented in a two-mode network visualization. As shown in Figure \ref{fig:example_introduction}, each event is a time-stamped \textit{directed hyperedge}, originating from a set of nodes of one type and pointing to a set of nodes of another type. However, since we have sets of authors linked to sets of cited papers in a single event, we also get relations between authors (co-authorship) and between papers (being co-cited). This allows for the computation of endogenous covariates that are functions of previously occurred relations of the type: (i) paper-cites-paper, (ii) author-cites-paper, (iii) author-cites-author, (iv) author-publishes-paper.}

There also exist relational hyperevents with a sender-receiver sets structure. For example, in a \textit{meeting}\index{meeting}, participants interact together at the same time without a sender-receiver distinction. In such cases, we can describe the event as an \textit{undirected hyperevent}\index{hyperevent!undirected hyperevent} by setting \(J_m = \emptyset\) and including all participants in \(I_m\). We denote the maximum size allowed for a meeting as \(w\). Mathematically, undirected hyperevents represent \textit{undirected hyperedges}\index{hyperedge!undirected hyperedge}, which are subsets of the vertex set \(V\).

%----------------------------------------------------------------------------%
\subsection{Relational Hyperevent Model}
\label{subsec_what_is_relational_hyperenetwork}
%----------------------------------------------------------------------------%

% Point process, counting process, and filtration
The event sequence in Equation~\eqref{eq_hyperevent_sequence} is treated as a realization of a \textit{marked point process}\index{marked point process}. In this process, the \textit{points}\index{marked point process!point} represent the times \(t\) when hyperevents occur, and the \textit{marks}\index{marked point process!mark} specify the senders \(I\) and receivers \(J\) involved in each event. Related to this point process is a \textit{counting process}\index{counting process} \(N(t, I, J)\), which records the number of hyperevents from group \(I\) to group \(J\) observed up to time \(t\). The dynamics of the counting process can be modeled using both exogenous and endogenous features. Here, \textit{exogenous}\index{exogenous} refers to factors external to the temporal hypergraph, while \textit{endogenous}\index{endogenous} refers to factors arising from within it. To capture these dynamics, we incorporate the \textit{event history} at time \(t\) in the sub-\(\sigma\)-algebra \(\mathcal{H}_{t} = \sigma \left( \{ (t_m, I_m, J_m) \mid t_m \leq t \} \right)\). If exogenous information is available -- {such as author score and paper impact in the synthetic example} -- it should be included in the \textit{filtration}\index{filtration} \(\mathbb{H} = \{\mathcal{H}_{t}\}_{t \ge 0}\).

% Counting process model for relational events
Under the assumptions that make \(N(\cdot, I, J)\) a right-continuous sub-martingale, adapted to the filtration \(\mathbb{H}\), we can apply the \textit{Doob-Meyer theorem}\index{Doob-Meyer theorem} \citep{perry2013point} to decompose {the counting process} into two parts: a \textit{noisy}\index{noisy} component \(M(\cdot, I, J)\), and a \textit{predictable}\index{predictable} component, the \textit{cumulative intensity process}\index{cumulative intensity process} \(\Lambda(\cdot, I, J)\). If it exists, the \textit{intensity function}\index{intensity function} is defined as
\(\lambda(t, I, J) = \dfrac{\partial \Lambda(t, I, J)}{\partial t}\).  
This function describes the \textit{instantaneous rate} at which the hyperevent \((t, I, J)\) occurs. We decompose this rate as follows:
\begin{equation}\label{eq_model_formulation}
    \lambda(t, I, J; \mathcal{H}_{t^-}) = W(t, I, J) \cdot \lambda_0(t) \cdot \exp{\left\{ f\left(\bm{x}(t, I, J), t\right) \right\}},
\end{equation}
where $\mathcal{H}_{t^-} = \sigma \left( \{ (t_m, I_m, J_m) \mid t_m < t \} \right)$. {In this paper, we use both the term rate and the term \textit{hazard} to indicate \(\lambda(t, I, J; \mathcal{H}_{t^-}) \)}.
This formulation is consistent with existing relational event and hyperevent models \citep{bianchi2024relational}. The term \(W(t, I, J)\) is the \textit{risk indicator}\index{risk indicator}, which determines whether the pair \((I, J)\) is at risk of a hyperevent at time \(t\). The term \(\lambda_0(t)\) is the \textit{baseline hazard}\index{baseline hazard}, representing the baseline rate of event occurrence, which may vary over time. The main focus of this paper is the \textit{edge-specific contribution function}\index{contribution function}, \(f\left(\bm{x}(t, I, J), t\right)\), which captures the effect of covariates on the event rate. In the next section, we discuss its specification and interpretation in detail.

%----------------------------------------------------------------------------%
\subsection{Effects in Relational Hyperevent Modeling}
\label{sec_effects_relational_hyperevent_modeling}
%----------------------------------------------------------------------------%

% Covariate evaluation
The contribution function examines how edge-specific \textit{covariates}\index{covariate} influence the event occurrence rate. Although the true multivariate \textit{covariate process}\index{covariate!covariate process} \(X(t, I, J)\) is unobserved, we can observe \(p\)-dimensional samples through the vector of \textit{measured covariates}\index{covariate!measured covariate} \(\bm{x}(t, I, J)\). The vector \(\bm{x}(t, I, J)\) may include \textit{exogenous covariates}\index{covariate!exogenous covariate}, such as attributes of the senders and/or receivers. For example, in a bibliometric hyper-relational graph of authors and papers, one might use aggregate statistics related to authors' demographics, academic titles, or institutional affiliations as exogenous covariates. 
{Figure \ref{fig:example_introduction} shows that each author has an attribute called \textit{Author Score} (S), while each paper has a \textit{Paper Impact} (I). To explain the rate of publication events, we use the \textit{average} author score of the authors and the \textit{total} paper impact of the referenced papers. This is an example of an exogenous hyperevent covariate, derived by aggregating information about the members participating in the hyperevent.}

{The vector \(\bm{x}(t, I, J)\) may also include}
\textit{endogenous covariates}\index{covariate!endogenous covariate}. In particular, we highlight a key aspect of event history modeling known as \textit{subset repetition}\index{subset repetition} \citep{lerner2024relational}. 
% subset repetition
Subset repetition allows for flexible evaluation of social structures based on past events, counting interactions between subgroups of participants over time. To define subset repetition, we begin by quantifying the frequency of interactions between two sets of participants using the \textit{activity operator}\index{activity operator}:
\begin{equation}\label{eq_action}
    \textup{activity}(t, I, J) = \sum_{t_m < t} \mathbbm{1}_{\{I \subseteq I_m \cap J \subseteq J_m\}},
\end{equation}
where \(\text{activity}(t, I, J)\) counts the number of past events in which all nodes in \(I\) (possibly with additional senders) jointly interacted with all nodes in \(J\) (possibly with additional receivers). Subset repetition\index{subset repetition} of order \((\rho, \ell)\) is then defined as the average number of sender-to-receiver interactions for all subsets of senders of size \(\rho\) and receivers of size \(\ell\):
\begin{equation}\label{eq_subrepetition}
    \textup{subrep}^{\rho, \ell}(t, I, J) = \sum_{(I', J') \in \binom{I}{\rho} \times \binom{J}{\ell}} \dfrac{\text{activity}(t, I', J')}{\binom{|I|}{\rho} \times \binom{|J|}{\ell}},
\end{equation}
where \(\binom{H}{\rho}\) denotes the set of all subsets of \(H\) of size \(\rho\), and \(\binom{|H|}{\rho}\) is the number of such subsets. In this work, we use the repeated interaction of sender and receiver subgroups -- captured through subset repetition -- as one of the key components for explaining hyperevent occurrences.

% Covariate effect
RHEM formulations in the current state of the art typically rely on the linearity assumption, which assumes that the contribution function is time-invariant and linear in the covariate values. A \textit{linear model}\index{linearity assumption!linear model} is expressed as \(f^{\text{LE}}\left(\bm{x}(t, I, J), t\right) = \bm{\theta}^\top \bm{x}(t, I, J),\) where \(\bm{\theta}\) is a vector of unknown coefficients expressing the time-invariant linear effect (LE) of covariates \(\bm{x}(t, I, J)\). 
{Consider the example in Figure~\ref{fig:example_introduction}, where we examine how the average author score and the total referenced papers' impact influence publication events. Assuming a linear effect for these covariates implies that their influence on the event rate remains constant between 1950 and 2021, i.e., the effect is homogeneous over time. It also imposes monotonicity: if the coefficient is positive, the contribution increases as more impactful papers are added to the reference list and never reaches saturation.}
We aim to relax this assumption by allowing each covariate to contribute in a non-linear way, with effects that may also vary over time. In the relational event modeling literature, more flexible specifications -- such as time-varying or non-linear effects -- have already been explored \citep{bauer2022smooth,boschi2023smooth,filippi2024stochastic, kamalabad2023point,uzaheta2024modeling}. Our goal is to extend these approaches to the relational hyperevent modeling framework, allowing covariate effects that are \textit{jointly} non-linear and time-varying. We first introduce the mathematical formulations for time-varying effects and non-linear effects, followed by the jointly time-varying non-linear effect\index{effect!jointly time-varying non-linear effect} (TVNLE).

% Effect Notation
For the remainder of this modeling section, we focus on the contribution of a single covariate \( x(t, I, J) \in \bm{x}(t, I, J) \) to the log-rate, denoted by \( f^\square\left(x(t, I, J), t\right) \). For simplicity, we sometimes omit the dependence on \( x(t, I, J) \) and \( t \), writing \( f^\square \) instead. The symbol \(\square\) indicates whether the contribution is time-varying (\(\square = \text{TVE}\)), non-linear (\(\square = \text{NLE}\)), or jointly time-varying and non-linear (\(\square = \text{TVNLE}\)). Under the \textit{additivity assumption}, the overall contribution function is expressed as \(f\left(\bm{x}(t, I, J), t\right) = \sum_{k=1}^p f^{(k), \square}\left(x^{(k)}(t, I, J), t\right)\), where \( x^{(k)}(t, I, J) \) is the \(k\)-th covariate in the vector \( \bm{x}(t, I, J) \), and \( f^{(k), \square} \) is its corresponding contribution function. Each \( f^{(k), \square} \) is assumed to be a \textit{smooth function}\index{smooth function} {or a \textit{tensor smooth}}.

\begin{figure}
    \begin{minipage}{0.45\textwidth}
        \centering
        a) \\
        \includegraphics[width=\linewidth]{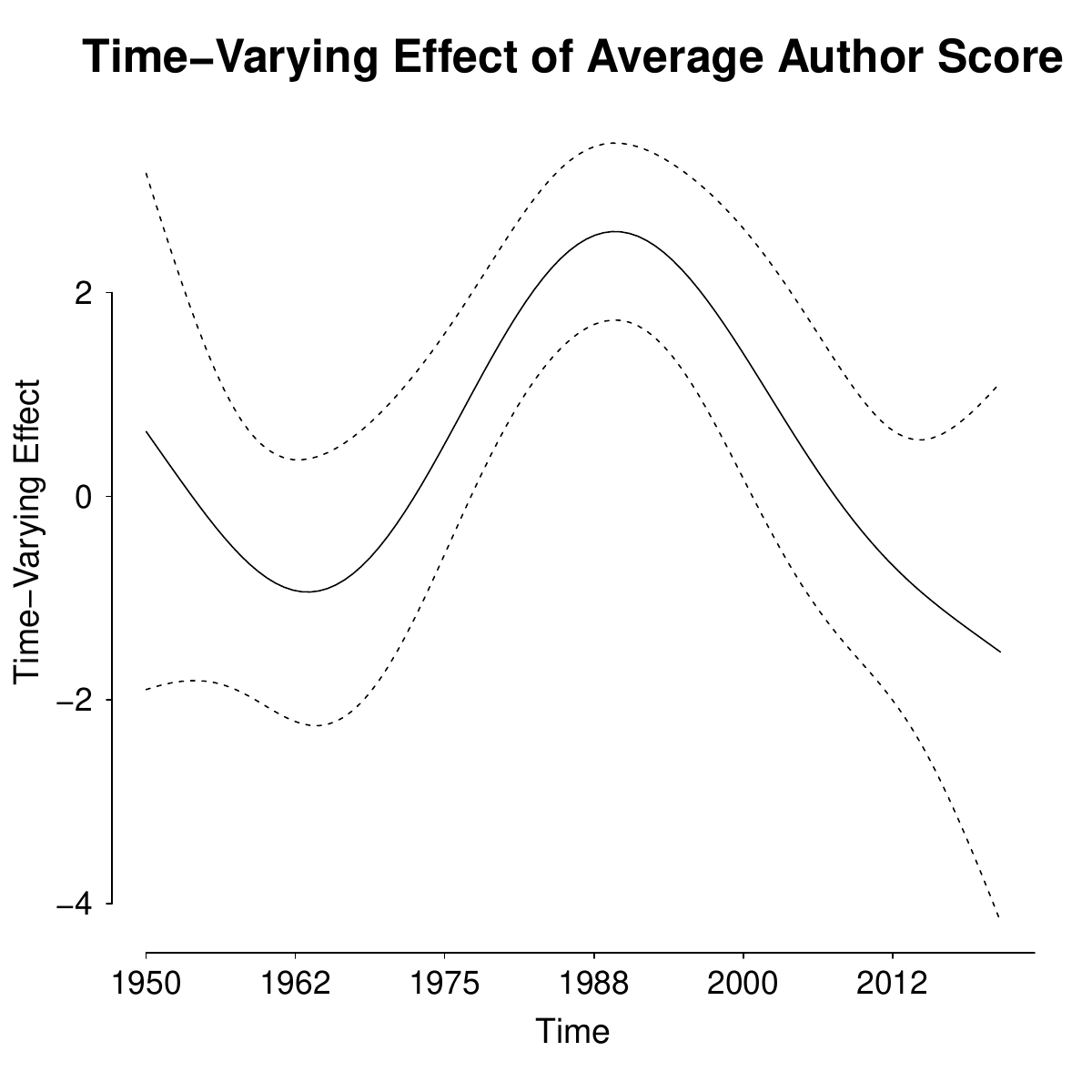}
    \end{minipage}
    \begin{minipage}{0.45\textwidth}
        \centering
        b) \\
        \includegraphics[width=\linewidth]{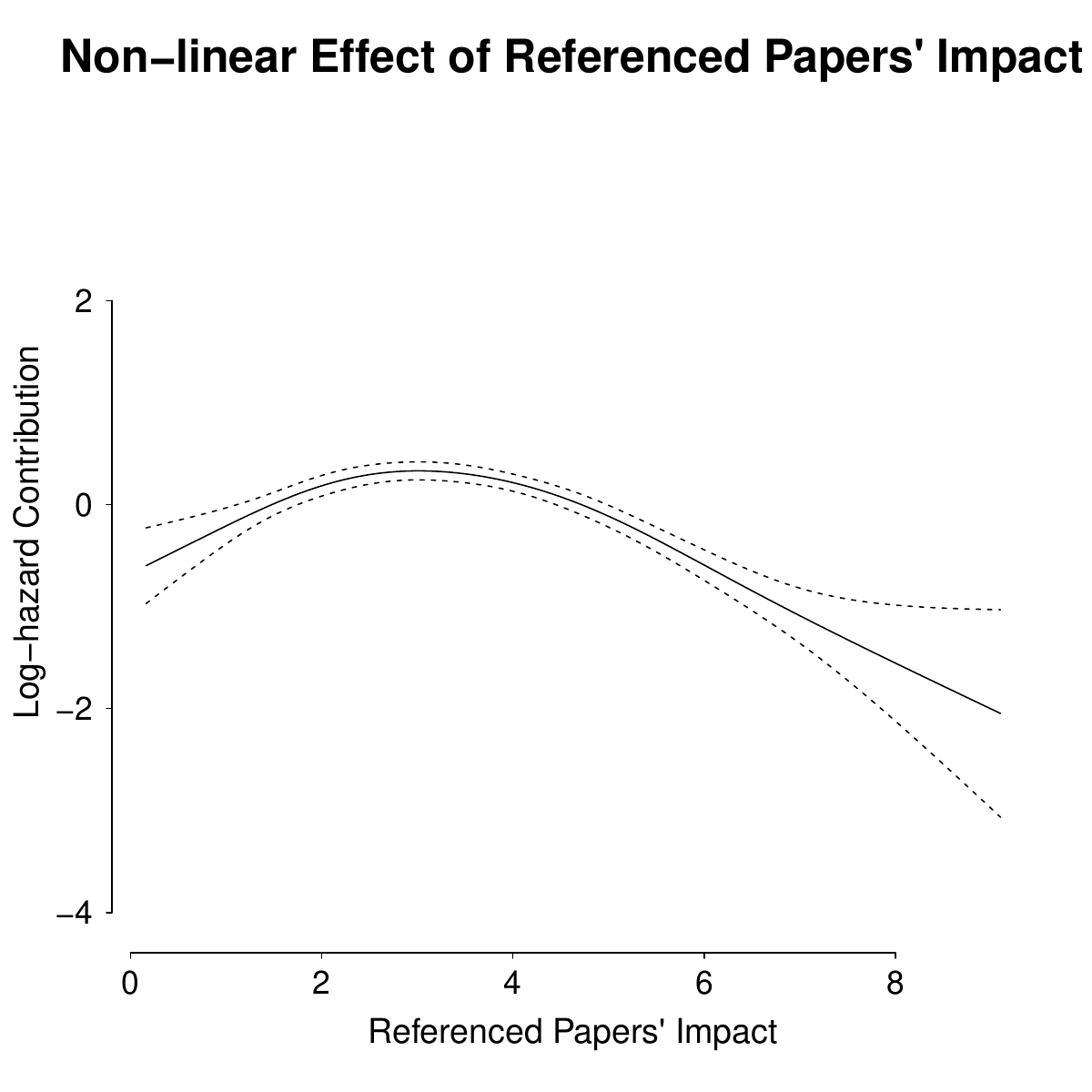}
    \end{minipage}
    {
    \caption{\label{fig:example_time_varying}{\bf Two hypothetical effects in the synthetic motivating example.}
    a) The solid curve represents the estimated time‐varying coefficient $\beta(t)$ of the Average Author Score (AAS), entering the model through the linear effect $\beta(t)\, x^{\text{AAS}}$. Together, these determine the contribution to the log-hazard. During the 1950s, authors with a high average score tended to publish less frequently. Toward the late 1980s this pattern reversed, with high-scoring authors publishing more often, although this tendency subsequently weakened. 
    b) The solid curve represents the contribution $f(x^{\text{RPI}})$ of the total Referenced Papers’ Impact (RPI) to the log-hazard. Initially, higher total RPI increases the likelihood of a publication event. However, when the total impact becomes too large, the contribution decreases, indicating a reduced likelihood of publication.}
    }
\end{figure}

% Time-varying effect 
\paragraph{Time-varying effect.}
The contribution of a covariate with a time-varying effect can be expressed as:
\begin{equation}\label{eq_time_varying_effect}
    f^{\text{TVE}}\left(x(t, I, J), t\right) = {\gamma(t)} \cdot x(t, I, J), \quad
    {\gamma(t)} = \sum_{l=1}^L \alpha_{l} a_{l}(t), 
\end{equation}
where {\(\gamma(t)\)} is a \textit{smooth function of time}\index{smooth function!smooth function of time} evaluated at \(t\) and representing the TVE of \(x(t, I, J)\). In its simplest form, {\(\gamma(t)\)} is represented as a linear combination of \(L\) non-linear basis functions of time \({a_{l}(t)}\), each weighted by a coefficient \(\alpha_l\). Note that \(f^{\text{TVE}}\) is linear with respect to the covariate value. Thus, \(f^{\text{TVE}}\) is the product of a smooth function of time evaluated at \(t\) and the covariate value at time \(t\). We generally refer to smooth function of time \(t\) as \(f_t\). 
{The scientific community’s perception of author scores may change. Although the score is synthetically generated, it reflects real situations in which features of scientific authors are viewed differently over time, for example due to new tools available to the community. In practice, this Average Author Score (AAS) does not contribute homogeneously to the log-occurrence rate but instead has a time-varying effect. According to Equation \eqref{eq_time_varying_effect}, this effect is a smooth function \({\gamma(t)}\). Its contribution to the log-hazard is obtained by multiplying the value of this function, at each time point, by the corresponding value of the covariate. An estimate of AAS's effect is shown in Figure \ref{fig:example_time_varying}a. Details of the estimation procedure are provided in Section \ref{sec_inference_procedures}.}

%  Non-linear effect
\paragraph{Non-linear effect.}
Non-linear effects allow the impact of a covariate to vary across different levels. For example, a covariate’s effect may increase up to a certain threshold, then plateau or even decrease beyond that point. The contribution of a covariate with a non-linear effect can be expressed as:
\begin{equation}\label{eq_non_linear_effect}
    f^{\text{NLE}}\left(x(t, I, J)\right) = \sum_{q=1}^Q \beta_q b_q\left(x(t, I, J)\right).
\end{equation}
Here, \(f^{\text{NLE}}\) is a \textit{smooth function of the covariate}\index{smooth function!smooth function of covariate} evaluated at time \(t\). In its simplest form, \(f^{\text{NLE}}\) is represented as a linear combination of \(Q\) non-linear basis functions \(b_q\), each weighted by a coefficient \(\beta_q\). We generally refer to smooth function of covariate \(x\) as \(f_x\). {It is plausible to suspect that citing high-impact papers may increase the likelihood of a paper being published. However, this effect may not be linear and could reach saturation. To investigate this, we model the non-linear influence of the total Referenced Papers' Impact (RPI) on the log-hazard, as shown in Figure \ref{fig:example_time_varying} b). As it is discussed in Section \ref{sec_inference_procedures}, estimated non-linear functions should be interpreted based on their relative changes and trends and \textit{not} on their sign. The plot suggests that while an increasing total RPI initially raises the probability of publication, this effect reaches a peak, after which citing additional highly impactful papers appears to have diminishing or even counterproductive effects.}

\begin{figure}
    \centering
    \begin{minipage}{0.45\textwidth}
        \centering
        a) \\
        \includegraphics[width=\linewidth]{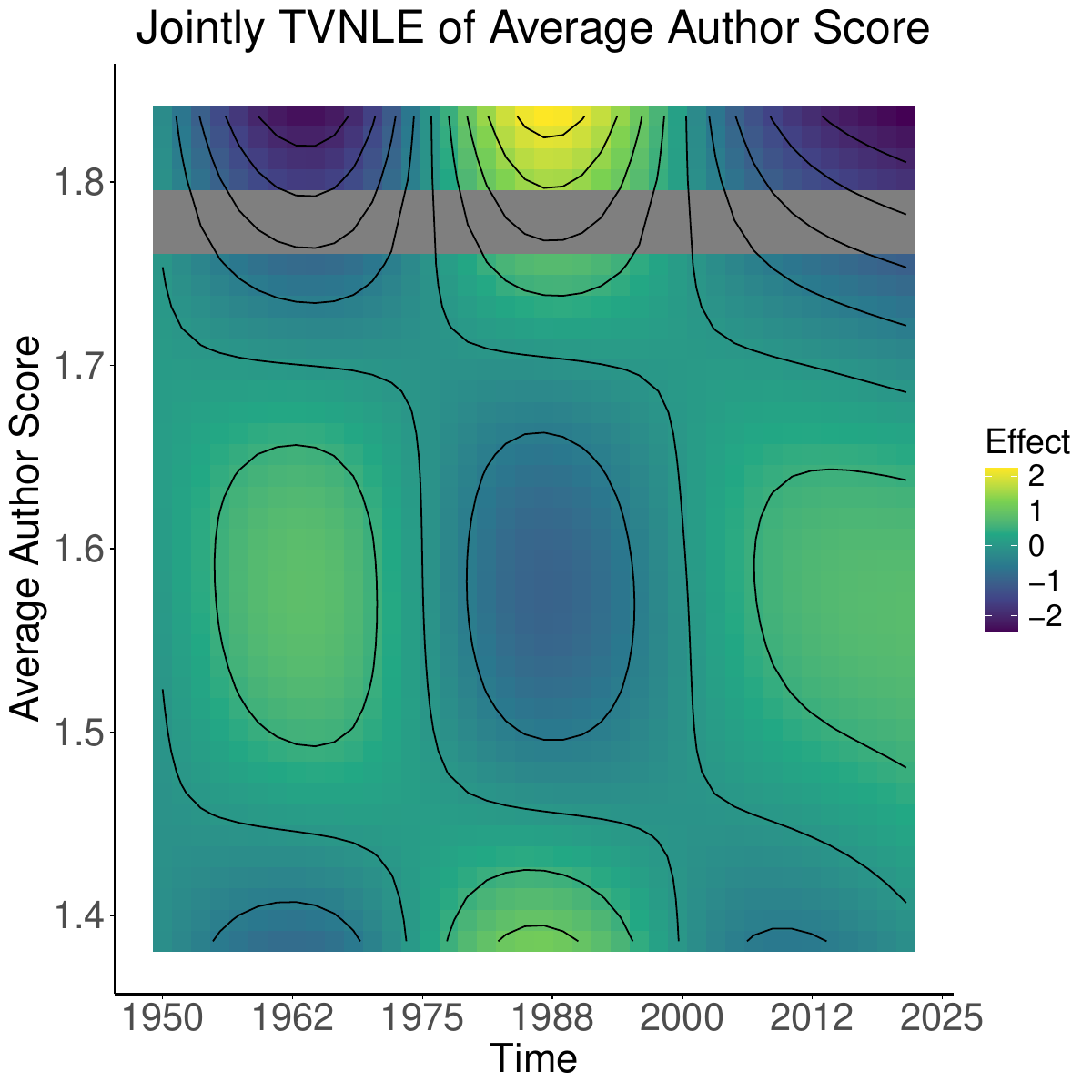}
    \end{minipage}
    \begin{minipage}{0.45\textwidth}
        \centering
        b) \\
        \includegraphics[width=\linewidth]{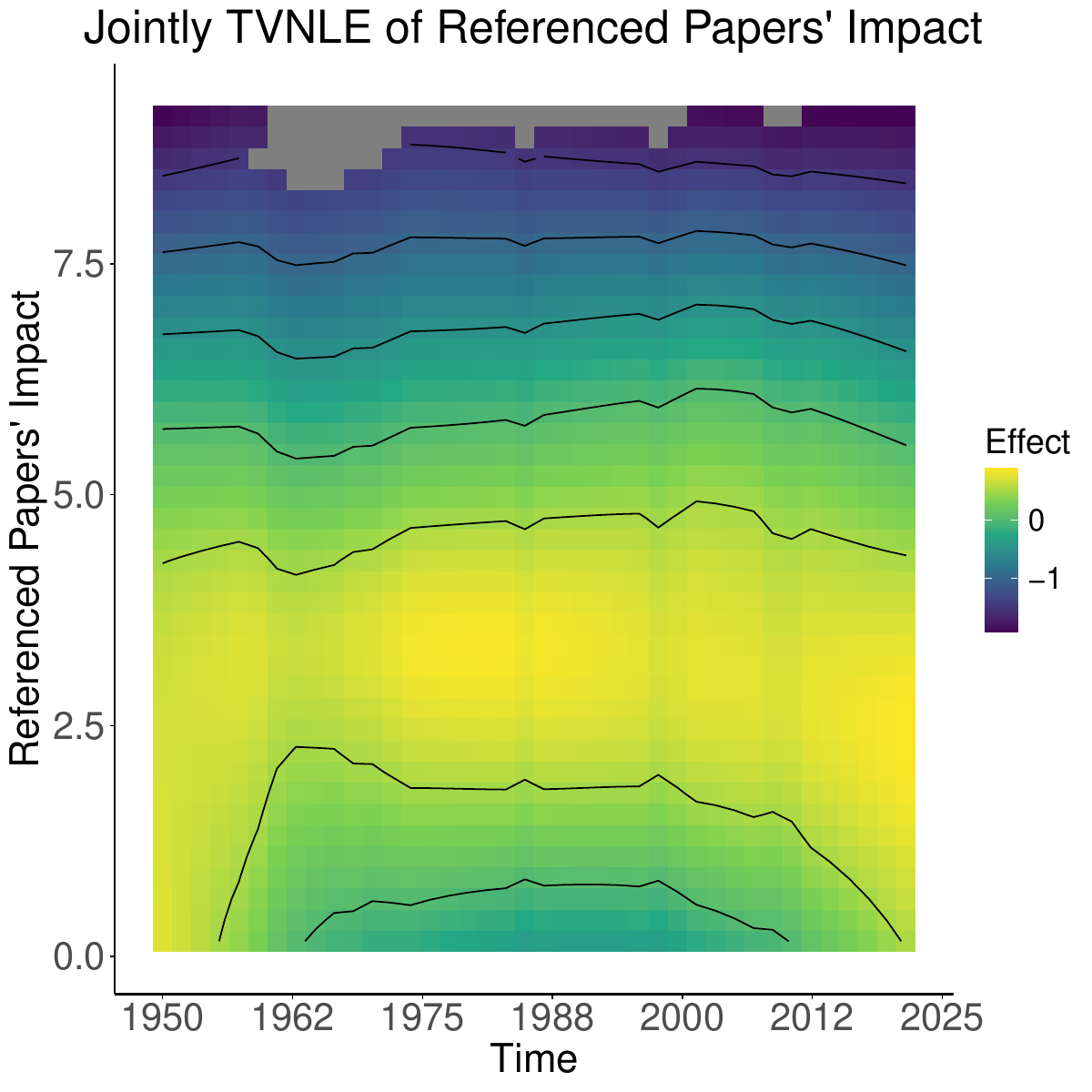}
    \end{minipage}
    {
    \caption{\label{fig:example_tvnle} {\bf Jointly time-varying non-linear effects of covariates in the synthetic motivating example.} 
    In both heatmaps, the x-axis represents time; fixing a point on the y-axis allows us to examine how the effect evolves over time. Since the y-axis represents covariate values, fixing a point on the x-axis shows the non-linear contribution of the covariate. Lighter (yellow) areas indicate a larger contribution to the log-hazard than darker (blue) areas. Gray points are values not observed in the data. Contour lines trace regions of constant effect; closer contours indicate faster changes.
    a) The heatmap shows that the contribution of the Average Author Score to the log-hazard varies non-monotonically over time and across covariate values. For higher values of AAS (approximately 1.7–1.85), the effect first decreases, then increases, and finally weakens again. This reflects the behavior observed in Figure \ref{fig:example_time_varying} a). At the beginning and end of the observation period (approximately 1950–1975 and 2000–2021), the trend is clearly non-monotonic, peaking around a value of 1.6. Between 1975 and 2000, the trend is inverted, favoring a higher average author score. b) The heatmap shows the contribution of the total Referenced Papers' Impact. Although the effect is non-linear (peaking around a value of 3), as in Figure~\ref{fig:example_time_varying} b), it remains relatively stable over time. This is visible from the horizontal contour lines: fixing a point on the y-axis reveals little temporal variation.}
}
\end{figure}

{
\paragraph{Shape-constraint non-linear effect.}
It is possible to add \textit{shape constraints} to the non-linear and time-varying effects in the RHEM by allowing the component smooth functions
$f^{(k), \square}$ to satisfy monotonicity or convexity constraints. The idea is based on the Shape Constrained Additive Models (SCAM) extension of generalized additive models \citep{pya2015shape}. They use a modified spline basis -- called \emph{SCOP-spline} -- in which the constraints are enforced
through a re-parameterization of the spline coefficients. Consider the spline representation:
\[
f^{\text{SCNLE}}(x(t, I, J)) = \sum_{q=1}^Q \delta_q B_q(x(t, I, J)),
\]
where $\{B_q\}$ are B-spline basis functions. A sufficient condition for $f'(x) \ge 0$ is that the coefficients form a non-decreasing sequence,
\(
\delta_1 \le \delta_2 \le \cdots \le \delta_Q.
\)
SCAMs impose this by reparameterizing
\(
\bm{\delta} = \Sigma \ \bm{\tilde\beta}\), where \( \bm{\tilde\beta} = (\beta_1,\exp(\beta_2),\ldots,\exp(\beta_Q))^\top,\) and $\Sigma$ is a cumulative-sum matrix ($\Sigma_{lq}=1$ for $l\ge q$ and $0$ otherwise).
With this parameterization, the constraints are automatically satisfied. 
}

% jointly time-varying non-linear effect 
\paragraph{Jointly time-varying non-liner effect.}
We aim to introduce a novel type of effect that, to the best of our knowledge, has not yet been explored in either the relational event or relational hyperevent modeling frameworks. Specifically, we propose using tensor products as smooth functions of several variables \citep{wood2017generalized} -- particularly time and covariate -- to model jointly time-varying non-linear effect\index{effect!time-varying non-linear effect} of covariates on event occurrence. The contribution of a covariate with a TVNLE can be expressed as:
\begin{equation}\label{eq_tvnle}
    f^{\text{TVNLE}}\left(x(t, I, J), t\right) = \sum_{q=1}^Q \left( \sum_{l = 1}^L \alpha_{ql} a_{l}(t) \right) b_q\left(x(t, I, J)\right).
\end{equation}
by allowing the coefficients $\beta_{q}$ in Equation~\eqref{eq_non_linear_effect} to vary smoothly in time, expressing them as smooth functions of time as in Equation~\eqref{eq_time_varying_effect}. We generally refer to a smooth function of time \(t\) and covariate \(x\) as \(f_{tx}\). 

{
Consider again the two previously introduced effects and their associated covariates. We have seen that the Average Author Score exhibits a time-varying effect, but it may also display a non-monotonic pattern, similar to the total Referenced Papers' Impact, where increases in the covariate can eventually become counterproductive for publication activity. Likewise, the influence of RPI may itself have evolved over time. To accommodate these possibilities, we relax the assumption of a smooth effect in only one dimension and instead allow for a jointly non-linear time-varying effect. Figure~\ref{fig:example_time_varying} shows the corresponding estimates. These estimates illustrate how the effect varies across both covariate values and time. In the plots, time is displayed on the x-axis; thus, to examine the temporal evolution of the effect, we fix a point on the y-axis. For higher AAS (approximately 1.7-1.85), the pattern in Figure~\ref{fig:example_time_varying} a) becomes evident: the effect first decreases, then increases, and finally weakens again. Conversely, since the y-axis represents covariate values, fixing a point on the x-axis allows us to observe the non-linear contribution of the covariate. At the beginning and end of the observation period (approximately 1950-1975 and 2000-2021), the trend is clearly non-monotonic, reaching a maximum for values around 1.6. Between 1975 and 2000, the trend is inverted, first decreasing and then increasing, favoring higher average author scores. For the AAS, including this jointly time-varying non-linear effect is crucial. In contrast, the total RPI appears largely stable over time. The non-linear pattern shown in Figure~\ref{fig:example_time_varying} b) can be observed by fixing any point on the x-axis and examining the variation across covariate values, showing a peak around a value of 3. However, when fixing a point on the y-axis, the effect remains fairly homogeneous over time.}

There are several options for the choice of spline basis functions in Equations~\eqref{eq_time_varying_effect}, \eqref{eq_non_linear_effect}, and \eqref{eq_tvnle}. In this paper, we focus on \textit{thin plate regression splines}\index{spline!thin plate regression splines} (TPRS) {and tensor product smooths, both implemented} in the \texttt{mgcv} package in \texttt{R} \citep{wood2017generalized}. While a comprehensive comparison of alternative spline types is beyond the scope of this work, we refer interested readers to \citet{wood2017generalized} for a detailed discussion. Our primary aim is indeed to underline the importance of incorporating flexible effect specifications in the modeling of relational hyperevents. 

% Mathematical considerations about the model
\paragraph{Mathematical considerations about the model.}\label{subsec_math_considerations}

TPRS are constructed by transforming and truncating the basis functions obtained from a \textit{thin plate spline}\index{spline!thin plate spline} (TPS) smoothing problem \citep{wood2003thin}. Within the TPS bases, two functions span the subspace of completely smooth terms \citep{wood2017generalized}: one is constant over the input variable, and the other is linear, perfectly correlated with the input variable itself. When modeling a time-varying effect \({\gamma(t)}\), as in Equation~\eqref{eq_time_varying_effect}, these basis functions include a constant term of the form \(\alpha_{l} a_{l}(t) = \alpha_{l} a_{l}\), where \(a_{l}\) is constant in time. This term behaves like a time-invariant coefficient, representing the effect the covariate would have under a standard linear assumption. In the case of a non-linear effect, as in Equation~\eqref{eq_non_linear_effect}, the contribution \(f^{\text{NLE}}\) is expressed as a linear combination of basis functions of the covariate, \(\beta_{q} b_{q}(x)\). When the basis function \(b_{q}(x)\) is linear, the term \(\beta_{q} b_{q}(x) = \beta_{q} b_{q} \cdot x\) corresponds to a linear effect, up to a linear transformation of the covariate itself. This structure can also show that both TVE and NLE can be interpreted as restricted versions of TVNLE, defined in Equation~\eqref{eq_tvnle}. If only the constant basis terms in time are active, the effect becomes non-linear in the covariate but time-invariant. Conversely, if only the linear basis terms in the covariate are retained, the effect becomes time-varying but linear in the covariate.

%----------------------------------------------------------------------------%

\subsection{Inference Procedures}
\label{sec_inference_procedures}
%----------------------------------------------------------------------------%

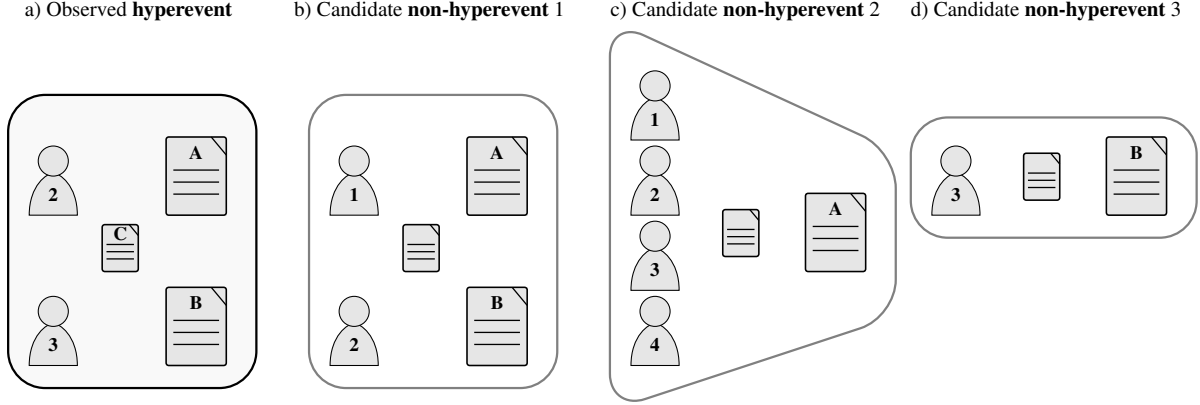
\begin{figure}[t]
\centering

\tikzset{
  pics/user/.style args={#1/#2}{  
    code={
      \draw[fill=#2, draw=black] 
        (-0.4,0) .. controls (-0.4,0.5) and (-0.1,0.7) .. (0,0.7)
        .. controls (0.1,0.7) and (0.4,0.5) .. (0.4,0) -- cycle;
      \node[font=\small] at (0,0.35) {\textbf{#1}};
      \filldraw[fill=#2, draw=black] (0,0.9) circle (0.25cm);
    }
  },
  pics/paper/.style args={#1/#2}{ 
    code={
      \fill[#2] (0,0) rectangle (1.0,1.3);
      \draw[line width=0.6pt, rounded corners=1pt] (0,0) rectangle (1.0,1.3);
      \foreach \y in {0.75,0.55,0.35} {
        \draw[line width=0.5pt] (0.12,\y) -- (0.88,\y);
      }
      \draw[line width=0.5pt] (0.75,1.3) -- (1.0,1.0);
      \node[font=\small] at (0.5,1.05) {\textbf{#1}};
    }
  }
}

\begin{tikzpicture}[every node/.style={scale=0.8}]

\begin{scope}[xshift=0cm]
  \coordinate (U2) at (0,0);
  \coordinate (U3) at (0,-2);
  \coordinate (P_A) at (1.5,0);
  \coordinate (P_B) at (1.5,-2);
  \coordinate (P_C) at (0.65,-0.75);

  \path[fill=gray!5, draw=black, line width=0.3mm, rounded corners=14pt]
    ($(U2)+(-0.6,1.6)$) --
    ($(P_A)+(1.2,1.6)$) --
    ($(P_B)+(1.2,-0.3)$) --
    ($(U3)+(-0.6,-0.3)$) -- cycle;

  \pic at (U2) {user={2/gray!20}};
  \pic at (U3) {user={3/gray!20}};
  \pic at (P_A) {paper={A/gray!20}};
  \pic at (P_B) {paper={B/gray!20}};
  \pic at (P_C) [scale=0.6] {paper={C/gray!20}};
  \node[above] at (1,2.5) {a) Observed \textbf{hyperevent}};
\end{scope}

% ---------------- First mini picture ----------------
\begin{scope}[xshift=4cm]
  \coordinate (U1) at (0,0);
  \coordinate (U2) at (0,-2);
  \coordinate (P_A) at (1.5,0);
  \coordinate (P_B) at (1.5,-2);
  \coordinate (P_C) at (0.65,-0.75);

  \path[draw=gray, line width=0.3mm, rounded corners=14pt]
    ($(U1)+(-0.6,1.6)$) --
    ($(P_A)+(1.2,1.6)$) --
    ($(P_B)+(1.2,-0.3)$) --
    ($(U2)+(-0.6,-0.3)$) -- cycle;

  \pic at (U1) {user={1/gray!20}};
  \pic at (U2) {user={2/gray!20}};
  \pic at (P_A) {paper={A/gray!20}};
  \pic at (P_B) {paper={B/gray!20}};
  \pic at (P_C) [scale=0.6] {paper={ /gray!20}};

  \node[above] at (1,2.5) {b) Candidate \textbf{non-hyperevent} 1};
\end{scope}

% ---------------- Second mini picture ----------------
\begin{scope}[xshift=8cm]
  \coordinate (U1) at (0,1);
  \coordinate (U2) at (0,0);
  \coordinate (U3) at (0,-1);
  \coordinate (U4) at (0,-2);
  \coordinate (P_A) at (2,-0.75);
  \coordinate (P_C) at (0.9,-0.55);
  
  \path[draw=gray, line width=0.3mm, rounded corners=14pt]
    ($(U1)+(-0.6,1.6)$) --
    ($(P_A)+(1.2,1.6)$) --
    ($(P_A)+(1.2,-0.6)$) --
    ($(U4)+(-0.6,-0.6)$) -- cycle;

  \pic at (U1) {user={1/gray!20}};
  \pic at (U2) {user={2/gray!20}};
  \pic at (U3) {user={3/gray!20}};
  \pic at (U4) {user={4/gray!20}};
  \pic at (P_A) {paper={A/gray!20}};
  \pic at (P_C) [scale=0.6] {paper={ /gray!20}};
  \node[above] at (1.2,2.5) {c) Candidate \textbf{non-hyperevent} 2};
\end{scope}

% ---------------- Third mini picture ----------------
\begin{scope}[xshift=12cm]
  \coordinate (U3) at (0,0);
  \coordinate (P_B) at (2,0);
  \coordinate (P_C) at (0.9,0.2);
  
  \path[draw=gray, line width=0.3mm, rounded corners=14pt]
    ($(U3)+(-0.6,1.3)$) --
    ($(P_B)+(1.2,1.3)$) --
    ($(P_B)+(1.2,-0.3)$) --
    ($(U3)+(-0.6,-0.3)$) -- cycle;

  \pic at (U3) {user={3/gray!20}};
  \pic at (P_B) {paper={B/gray!20}};
  \pic at (P_C) [scale=0.6] {paper={ /gray!20}};
  \node[above] at (1.2,2.5) {d) Candidate \textbf{non-hyperevent} 3};
\end{scope}

\end{tikzpicture}
{
\caption{\label{fig:event_nonevents} \textbf{Observed hyperevent and candidate non-hyperevents in the synthetic motivating example.}
a) Observed hyperevent. This corresponds to the first hyperevent shown in Figure~\ref{fig:example_introduction} and listed in Table~\ref{tab:hyperevents}, where authors~2 and~3 publish article~C citing papers~A and~B. To compute the likelihood in Equation~\ref{eq_partial_likelihood}, a non-hyperevent must be sampled for comparison. 
Panels b), c), and d) illustrate possible candidates. In the empirical application, non-hyperevents are sampled to have the same cardinality as the observed hyperevent. Under this criterion, only the non-hyperevent in panel~b) is a valid candidate among those shown.}
}
\end{figure}

To implement the estimation procedure based on the data in Equation~\eqref{eq_hyperevent_sequence}, we use a sampled version of the \textit{partial likelihood}\index{likelihood!partial likelihood} \citep{lerner2020reliability}. Specifically, at each event time, we consider only two instances: the observed hyperevent that actually occurred, and a \textit{non-hyperevent}\index{non-hyperevent} -- an event that did not occur but could have -- randomly sampled from those at risk. {Consider the first hyperevent in Figure \ref{fig:example_introduction}. The observed publication event involves authors 2 and 3 citing papers A and B, as also reported in the left panel of Figure \ref{fig:event_nonevents}. This event represents just one of the 45 possible events that could have occurred at that time, given all potential collaborations among authors 1-4 and papers A and B. Other panels of Figure \ref{fig:event_nonevents} show examples of non-hyperevents.}

Conditioning each observation on the event history and the corresponding event time, the \textit{case-control partial likelihood}\index{likelihood!case-control partial likelihood} can be expressed as:
\begin{equation}\label{eq_partial_likelihood}  
    \begin{aligned}
        \mathcal{L}(\bm{\theta}) =& \prod_{m=1}^n \dfrac{\exp{\{ f(\bm{x}(t_m, I_m, J_m); \bm{\theta})\}}}{  
        \exp{\{ f(\bm{x}(t_m, I_m, J_m); \bm{\theta})\}} +  
        \exp{\{ f(\bm{x}(t_m, I_m^\ast, J_m^\ast); \bm{\theta})\}}}  \\
        =& \prod_{m=1}^n \text{logistic} \left[ f(\bm{x}(t_m, I_m, J_m); \bm{\theta}) - f(\bm{x}(t_m, I_m^\ast, J_m^\ast); \bm{\theta}) \right],
    \end{aligned}
\end{equation}
where \((I_m^\ast, J_m^\ast)\) denotes the sender and receiver sets of a non-hyperevent sampled at time \(t_m\). In our empirical application, the non-event is constrained to have the same sender and receiver cardinalities as the observed event, i.e., \((I_m^\ast, J_m^\ast) \in \binom{V^I_{t_m}}{|I_m|} \times \binom{V^J_{t_m}}{|J_m|}\). {Practically, this means that among the three non-hyperevents shown in Figure \ref{fig:event_nonevents}, only the first one would be eligible.} Here, \(\bm{\theta}\) denotes the vector containing all coefficients associated with the basis functions used to represent the smooth covariate effects, or, in the special case of a linear effect, the single coefficient associated with the covariate. {The idea behind the likelihood in Equation \ref{eq_partial_likelihood} is that, at each time point, an observed hyperevent is compared with a sampled non-hyperevent -- similar to comparing panels a) and b) of Figure \ref{fig:event_nonevents}. To assess whether a covariate is relevant, we examine how its value differs between hyperevents that actually occur and their corresponding non-hyperevents. This difference then serves as the explanatory variable in the logistic formulation introduced above. In this context for hyperevent in a) and non-hyperevent in b), we evaluate AAS and the total RPI.} Following \citet{boschi2023smooth}, Equation~\eqref{eq_partial_likelihood} corresponds to the likelihood of a degenerate logistic regression, where the response is fixed (equal to 1), the intercept is omitted, and the predictors are given by the difference in the contribution function evaluated between the observed event and the sampled non-event. As a result, we can fit our model {as a Generalized Linear Model (GLM) with a logit link function, rather than relying on traditional survival modeling techniques. When the linearity assumption is relaxed and replaced with an additivity assumption -- as done in Section~\ref{sec_effects_relational_hyperevent_modeling} -- the model becomes a \textit{Generalized Additive Model}\index{generalized additive model} (GAM). This framework allows for more flexible and complex effects, such as those illustrated in Figures~\ref{fig:example_time_varying} and~\ref{fig:example_tvnle}.}
 
{To reduce potential overfitting, especially when using smooth functions, we employ a \textit{penalized log-likelihood}\index{likelihood!penalized log-likelihood} as the objective function. The penalized log-likelihood is defined as the log of Equation \eqref{eq_partial_likelihood} minus a penalty term \(P\). This term is a weighted sum of individual penalties \(P^{(k)}\), each associated with one of the \(p\) smooth components. Covariates modeled linearly receive no penalty, i.e., their penalty is zero. Each penalty is weighted by a \textit{smoothing parameter}\index{smoothing parameter}, which controls the trade-off between model fit and smoothness. These smoothing parameters are collected in the vector \(\bm{\tau}\). Using \texttt{R} package \texttt{mgcv}, smoothing parameters can be estimated directly from the data using cross-validation, restricted maximum likelihood, or maximum likelihood.}

For a univariate smooth effect \(f^{\text{NLE}}\), also denoted by \(f_x\), a common penalty is \(P(f_x) = \displaystyle \int \left( \frac{\partial^2 f_x}{\partial x^2} \right)^2 dx\), which quantifies the \textit{wiggliness}\index{wiggliness} of the function. As \(\tau \to \infty\), the function \(f_x\) is increasingly constrained toward linearity, ultimately reducing to a straight line. More complex penalty expressions are used in the case of TPRS. However, expressing \(f_x\) as a linear combination of basis functions, as in Equation~\eqref{eq_non_linear_effect}, allows each penalty \(P_x(f_x)\) to be written as a quadratic form in the corresponding coefficient vector. {To control smoothness, SCAMs use a difference penalty on the (unconstrained) spline coefficients $\bm{\beta}$,
\(
P(\bm{\beta}) = \nu \lVert D \bm{\beta} \rVert^2,
\)
where $D$ is a first- or second-order finite-difference operator.}

When dealing with multivariate smooth functions -- and in particular, with functions of both time and covariate, as in our main case of interest -- we denote the TVNLE effect by \(f_{tx}\). The associated penalty can be expressed as:
\begin{equation*}
    P\left(f_{tx}\right) = \tau_t \int_x P(f_{t|x}) \, dx + \tau_x \int_t P(f_{x|t}) \, dt
\end{equation*}
where \(\tau_t\) and \(\tau_x\) control the the tradeoff between wiggliness in different directions and \(f_{t|x}\) is \(f_{tx}\) considered as function \(t\), holding \(x\) constant. If we refer to the previously mentioned \(P(f_x)\) used as an example, the penalty term in the multivariate case would be \(P(f_{tx}) =  \displaystyle \int_{t,x} \tau_t \left( \frac{\partial^2 f_{tx}}{\partial t^2} \right)^2 + \tau_x \left( \frac{\partial^2 f_{tx}}{\partial x^2} \right)^2 \, dtdx\) \citep{wood2017generalized}. 

{In practice, we use the \texttt{gam} function from the \texttt{mgcv} package in \texttt{R} \citep{wood2017generalized}, that automatically implements the penalization. To fit a TVE, we include the term \texttt{s(time, by=delta\_x)}, where \texttt{time} corresponds to the event times and \texttt{delta\_x} represents the difference between the covariate evaluated for the event and the non-event. This fits a smooth function of time multiplied by the covariate. An example of the estimate (solid), along with its confidence intervals (dashed), is shown in Figure \ref{fig:example_time_varying} a).}
{To fit a NLE, we include the term \texttt{s(X, by=I)}, where \texttt{X} is a matrix with the first column containing the covariate values for the event and the second column containing those for the non-event. \texttt{I} is a two-column matrix, with the first column filled with \(1\)s and the second with \(-1\)s. An example of an estimated NLE (solid), along with its confidence intervals (dashed), is shown in Figure \ref{fig:example_time_varying} b). The estimated function cannot be interpreted by its sign due to a centering constraint; only its relative change and trend are meaningful.}
{Finally, to estimate a TVNLE, we include the term \texttt{te(T, X, by=I)}, where \texttt{X} and \texttt{I} are defined above, and \texttt{T} is a matrix with both columns equal to the vector of event times. As with the NLE, centering constraints are applied. Therefore, the estimated function needs to be centered, so that its average at each time point is zero. Two examples of centered TVNLE are reported in Figure \ref{fig:example_tvnle}, with the gray squares indicating combinations of covariate and time values that not observed in the data. As before, the estimated effect cannot be interpreted by its sign, only by its relative change and trend. On contour lines, the effect remains constant; the closer the contour lines, the faster the effect changes.}

\paragraph{Avoiding overfitting.}
{Even though our relational event model allows for smooth, non-linear effects, the estimation procedure is designed to prevent these flexible components from overfitting the data. As we described abov, each smooth term is equipped with a penalty that determines how freely it can adapt to patterns in the covariates. Rather than setting this by hand, we let the data decide: the inference scheme treats these smooth components as random effects and selects their level of flexibility by maximizing the marginal likelihood. In practical terms, this means that overly complex curves are automatically penalized, while genuinely nonlinear structures are retained. As a result, the estimated model possesses the expressive power of non-linear effect without succumbing to the instability and overfitting often associated with highly flexible specifications.}

\subsection{Computational Cost}

{In this section, we have shown how to define and estimate different types of effects -- linear, time-varying, non-linear, shape-constraint non-linear and jointly time-varying non-linear. All inference procedures rely on optimizing the penalized log-likelihood of a logistic additive regression. For a model matrix of dimension \(D\), the main computational cost arises from inverting the Hessian matrix, which requires \(\mathcal{O}(D^3)\) operations. A covariate with a linear effect uses a single column in the model matrix, while time-varying and non-linear effects use \(L\) and \(Q\) columns, respectively. A jointly time-varying non-linear effect uses \(L \times Q\) elements, with the tensor product smooth for an observation \(m\) given by the Kronecker product of the marginal smooth evaluations. When all \(p\) covariates are fitted with the same type of effect, the computational cost increases from \(\mathcal{O}(p^3)\) to \(\mathcal{O}(p^3 L^3 Q^3)\) operations. Memory requirements follow the same principle, growing from \( O(n  p)\) for linear effects to \(O(np L Q)\) for smooth effects. Importantly, \(L\) and \(Q\) are constants and independent of network features.}

{The cost for computing hyperedge covariates $x(t,I,J)$ is not affected by the effects becoming non-linear and/or time-varying. \cite{lerner2024relational} demonstrate that the computation of hyperedge covariates can scale to networks with more than a million nodes and events. Notably, the logistic regression formulation reduces the computational cost of likelihood evaluation, making it linear in the number of events \citep{lerner2020reliability, boschi2023smooth}. Importantly, this computation does not depend on the size of the relational network, unlike full partial likelihood-based estimation procedures.}

    % SIMULATION STUDY
    \section{Simulation study} 
\label{sec_simulation_study}

\begin{figure}
    \centering
   
    \begin{minipage}{0.45\linewidth}
        \centering
        a) n = 5000 (Linear vs NLE Model)\\
        \includegraphics[width=\linewidth]{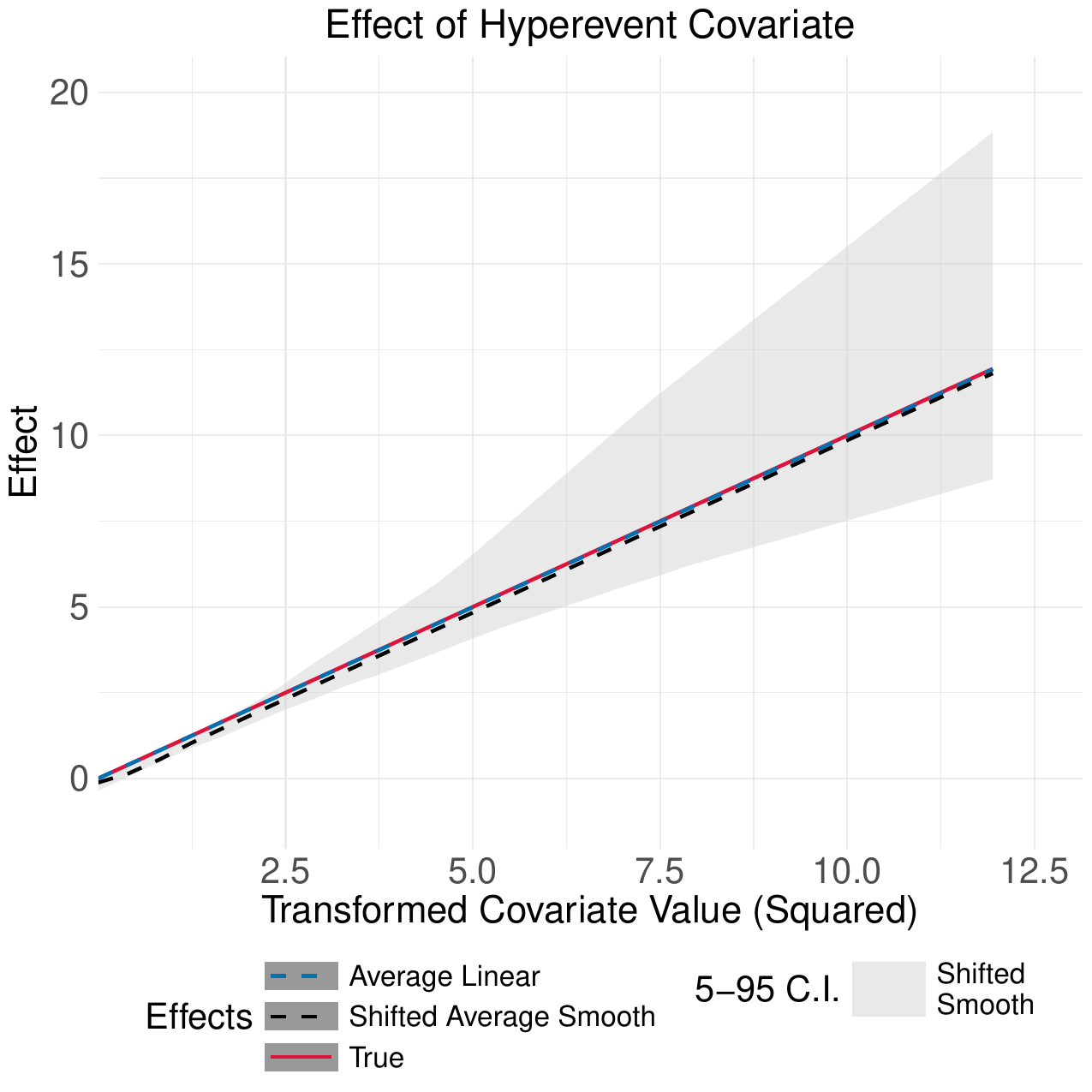}
    \end{minipage}
    \begin{minipage}{0.45\linewidth}
        \centering
        b) n = 10000 (Linear vs NLE Model) \\
        \includegraphics[width=\linewidth]{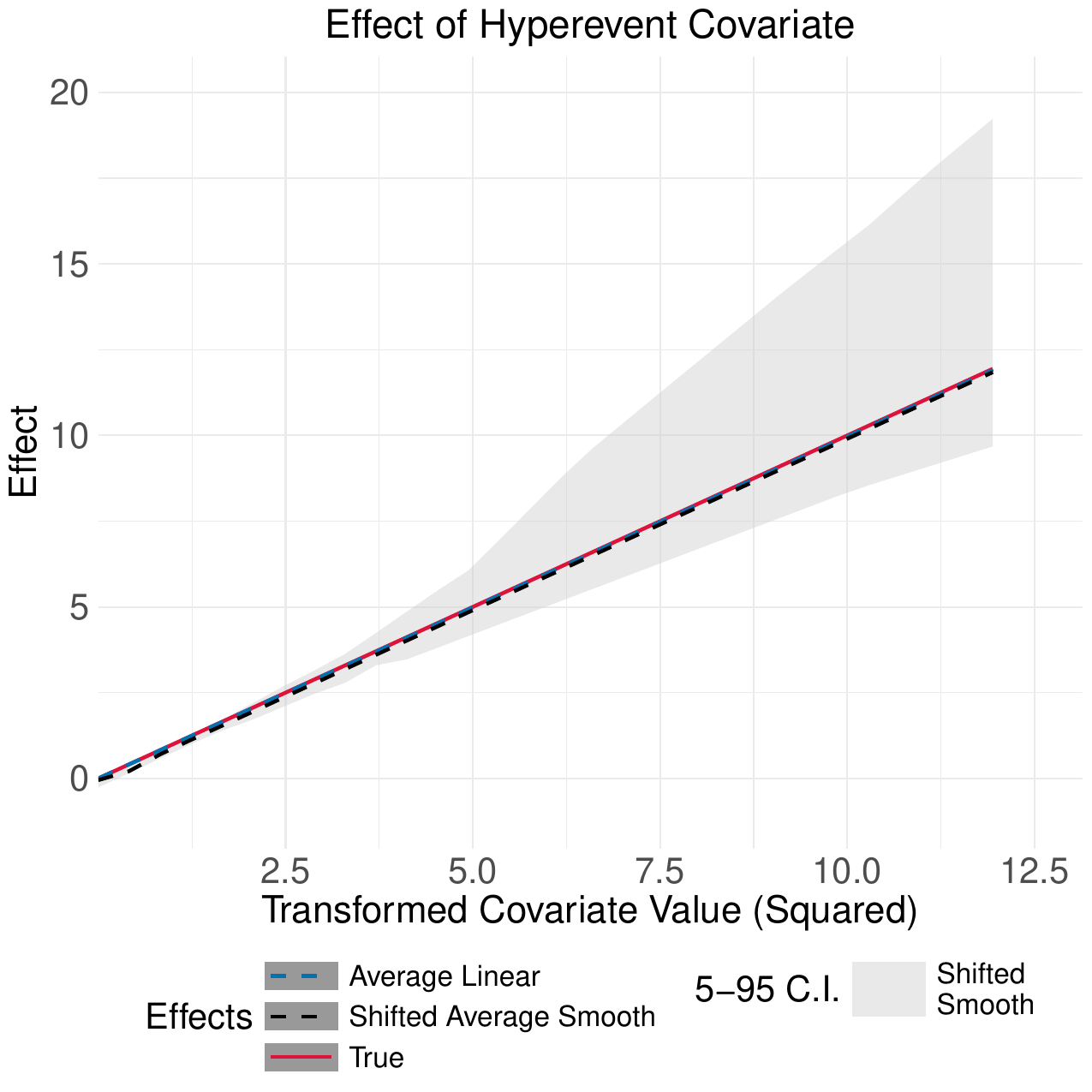}
    \end{minipage}
    \begin{minipage}{0.45\linewidth}
        \centering
        c) n = 10000 (TVNLE Model) \\
        \includegraphics[width=\linewidth]{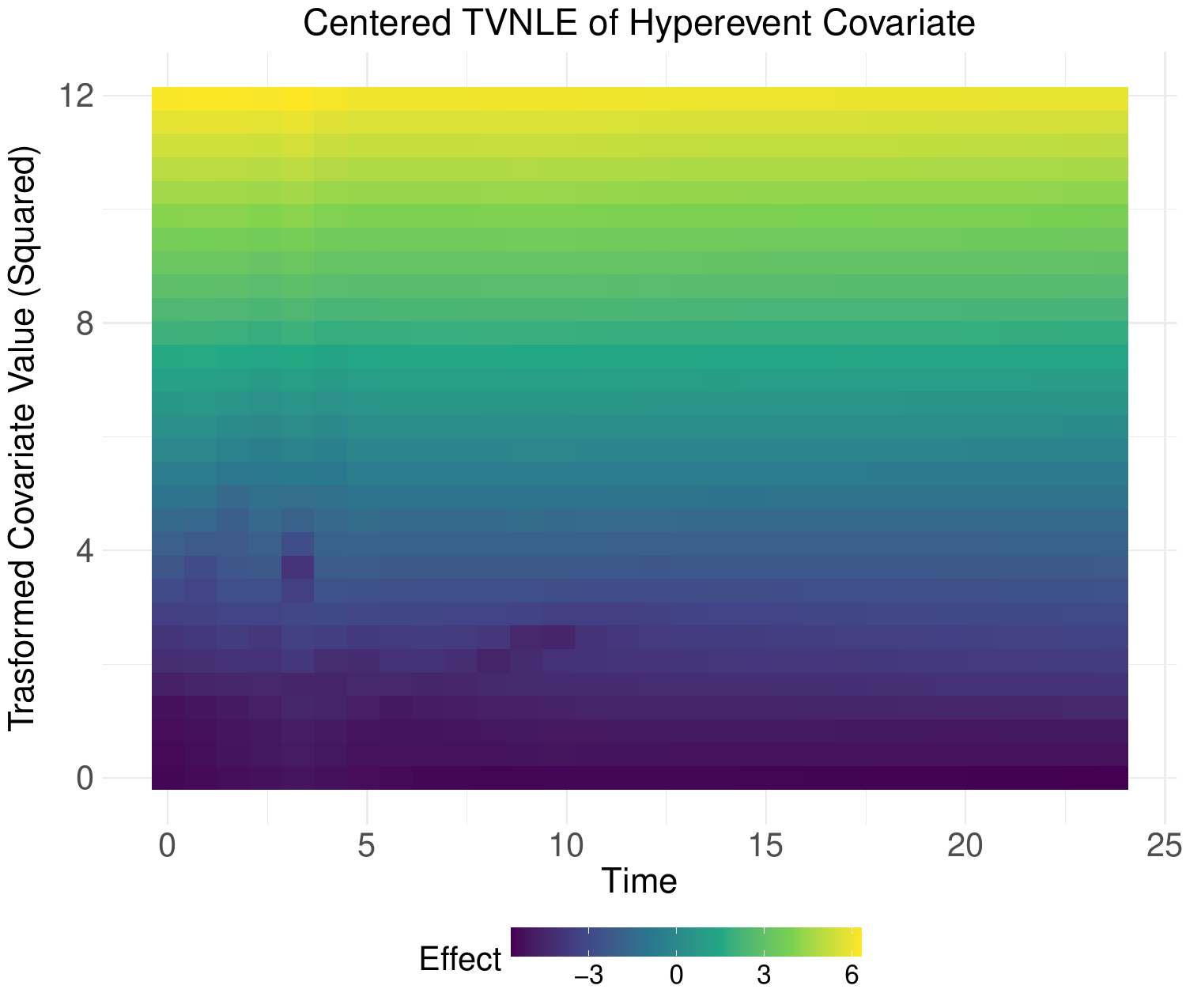}
    \end{minipage}
    \caption[Linear models are recovered by penalized non-linear models.]{\label{fig:simulation_study_linear}\textbf{Linear models are recovered by penalized non-linear models.} Panels a), b), and c) display results based on synthetic data generated according to the model in Equation~\ref{eq_RG1}, differing only in the number \(n\) of simulated events. Panels a) and b) show estimates from linear and non-linear models, while panel c) presents results from a model incorporating a joint time-varying and non-linear effect. 
    \emph{Top}. Across experiment replications, estimates are aggregated using inverse variance weighting\index{inverse variance weighting}. For the linear model, this yields a consensus slope referred to as the \enquote{consensus linear effect} (blue dashed line). For non-linear models, predicted effects are interpolated across the covariate domain, and aggregated pointwise using inverse variance weights, resulting in the \enquote{consensus non-linear effect} (black dashed line). Since non-linear estimates are identifiable only up to an additive constant, the consensus curve is manually shifted for alignment. Confidence intervals, derived from the 5th and 95th percentiles of the empirical interpolated estimates at each covariate value, are also shifted accordingly. Both linear and non-linear estimated effects are compared to the ground truth (red solid line). \emph{Bottom}. For the TVNLE model, estimates are aggregated using inverse variance weighting. The resulting smooth surface is then centered such that, at each time point, the average effect across the covariate domain is equal to \(0\).}
\end{figure}

To validate our approach and demonstrate its potential, we apply it to synthetic hyperevent data. 
We generate undirected hyperevents of the form \((t_m, I_m, J_m = \emptyset)\), representing meetings\index{meeting}. Each meeting involves a varying number of participants in \(I_m\), with event sizes ranging from 1 up to a predefined maximum \(w\), which defines the largest possible event (\(|I_m| \leq w \ \forall m\)). Events are generated from a known underlying model, as described in Equation~\eqref{eq_model_formulation}. We consider four explanatory variables. Two are endogenous covariates, represented by first- and second-order subset repetition covariates (as defined in Equation~\eqref{eq_subrepetition} with \(\rho = 1, 2\), respectively). One exogenous covariate corresponds to the average of a quantitative feature evaluated across all meeting participants (i.e., nodes in the hyperedge). This individual-level feature is drawn from a Gaussian distribution, with its mean and standard deviation defined as part of the simulation setup. Lastly, event size is included as an explanatory variable, modeled as a size penalty: it contributes negatively to the log-occurrence rate of the hyperevent. 

For each simulated hyperevent, we allow the corresponding non-observed meeting\index{non-hyperevent} to take any possible size from 1 to \( w \). This setup offers greater flexibility compared to the strategy adopted in our empirical application, where non-hyperevents are sampled to match the size of the observed hyperevent. In particular, allowing variation in non-event sizes enables us to estimate the effect of event size. However, this flexibility is feasible primarily because we are working with a relatively small hypergraph. As the number of actors increases, the number of possible hyperedges grows exponentially, making this approach computationally infeasible at larger scales.
 
Depending to the simulation setup, covariates effects are selected as LE, TVE, NLE, and TVNLE. This study was designed with four objectives in mind. First, we aim to demonstrate that when the true underlying effect is linear, both NLE and TVNLE specifications are flexible enough to recover linearity as a special case -- as discussed in Section~\ref{subsec_math_considerations}. Second, we highlight that when the true effect varies over time, fitting a linear model fails to capture the temporal dynamics accurately. Third, we show the potential of the proposed joint TVNLE estimation procedure in capturing complex covariate effects. Finally, we emphasize that using only time-varying or only non-linear models may be insufficient when the data-generating process involves both temporal and non-linear components. 

Each simulation scenario presented in this study is replicated \(100\) times. Since the synthetic datasets differ across replications, the observed ranges of both time and covariate values may vary accordingly. As a consequence, some predicted values may correspond to out-of-sample observations for models that were fitted on more limited ranges of the covariate or time. 

\subsection{Linear models are recovered by penalized non-linear models.}

{In this section, we simulate data from a rate that is linear in one of the covariates, $\bar{x}(I)^2$, but we fit it using a non-linear effect, $f(\bar{x}(I)^2)$. The aim is to assess whether the non-linear model overfits the effect. More specifically, the underlying true data-generating process has the following intensity function:}
\begin{equation}\label{eq_RG1}
    \log\lambda(t, I) = -\log\left(\sqrt{\text{subrep}^{1}(t, I)} \right) + \log\left(\sqrt{\text{subrep}^{2}(t, I)}\right) + \overline{x}(I)^2 - 0.5\cdot |I|.
\end{equation}

{Simulated events are undirected hyperevents, i.e., meetings involving multiple actors simultaneously, without distinguishing between senders and receivers.}. Consequently, the intensity function depends solely on the set \(I\) of participants. The terms \(\text{subrep}^{1}(t, I)\) and \(\text{subrep}^{2}(t, I)\) are endogenous covariates capturing first- and second-order subset repetition, respectively, as defined in Equation~\eqref{eq_subrepetition}. The variable \(\overline{x}(I)\) is an exogenous covariate computed as the average of the individual feature across all members of \(I\). Finally, \(|I|\) denotes the size of the meeting, included in the model with a negative contribution to penalize larger events. If \( \overline{x}(I)^2 \) is used as the input covariate -- instead of \( \overline{x}(I) \) -- its effect remains linear within the model specification, as the transformation is applied to the covariate before modeling. {Note that the coefficients for first- and second-order subset repetition are chosen with opposite signs. If both were positive, the simulation would become unstable due to the nature of this covariate.}

Figure~\ref{fig:simulation_study_linear} aggregates both linear and non-linear estimates from multiple replications using \textit{inverse-variance weighting}\index{inverse-variance weighting} \citep{hartung2011statistical}. 

Specifically, for the linear model, we aggregate the estimated slopes. For non-linear effects, we first interpolate the predicted effects over the covariate values, then perform point-wise aggregation, weighting each contribution inversely by its corresponding variance. Since estimated non-linear functions are interpretable only in terms of their trend, rather than their absolute sign or value, we manually shifted the consensus line (and relative confidence intervals) to facilitate direct comparison with the true and linear model estimates. As shown in Figure~\ref{fig:simulation_study_linear} a) and b), the flexible model successfully captures the true linear trend. Also, comparing results for different numbers of meetings \(n\), we can see that as \(n\) increases, there is a reduction in the variability of the non-linear estimates across different replications. 

Similarly, we interpolate joint time-varying non-linear effects using inverse-variance weighting. Furthermore, since TVNLE can only be identifiable up to the addition of an arbitrary function \(f_0(t)\), depending on time but not on the covariate, we center the value in such a way that they have zero-mean for each value of time. The fitted TVNLE models successfully recover a linear, time-invariant trend, as shown in Figure~\ref{fig:simulation_study_linear} c). Linearity can be seen by fixing a point on the $x$-axis (time $t$) and observing that the effect values vary monotonically with covariate values on the $y$-axis, increasing from bottom to top. Time-invariance is confirmed by fixing a point on the $y$-axis (covariate value) and noting that the effect remains essentially constant, from left to right, over time.

\begin{figure}
    \centering
    \begin{minipage}{0.45\linewidth}
        \centering
        a) Linear vs TVE Model \\
        \includegraphics[width=\linewidth]{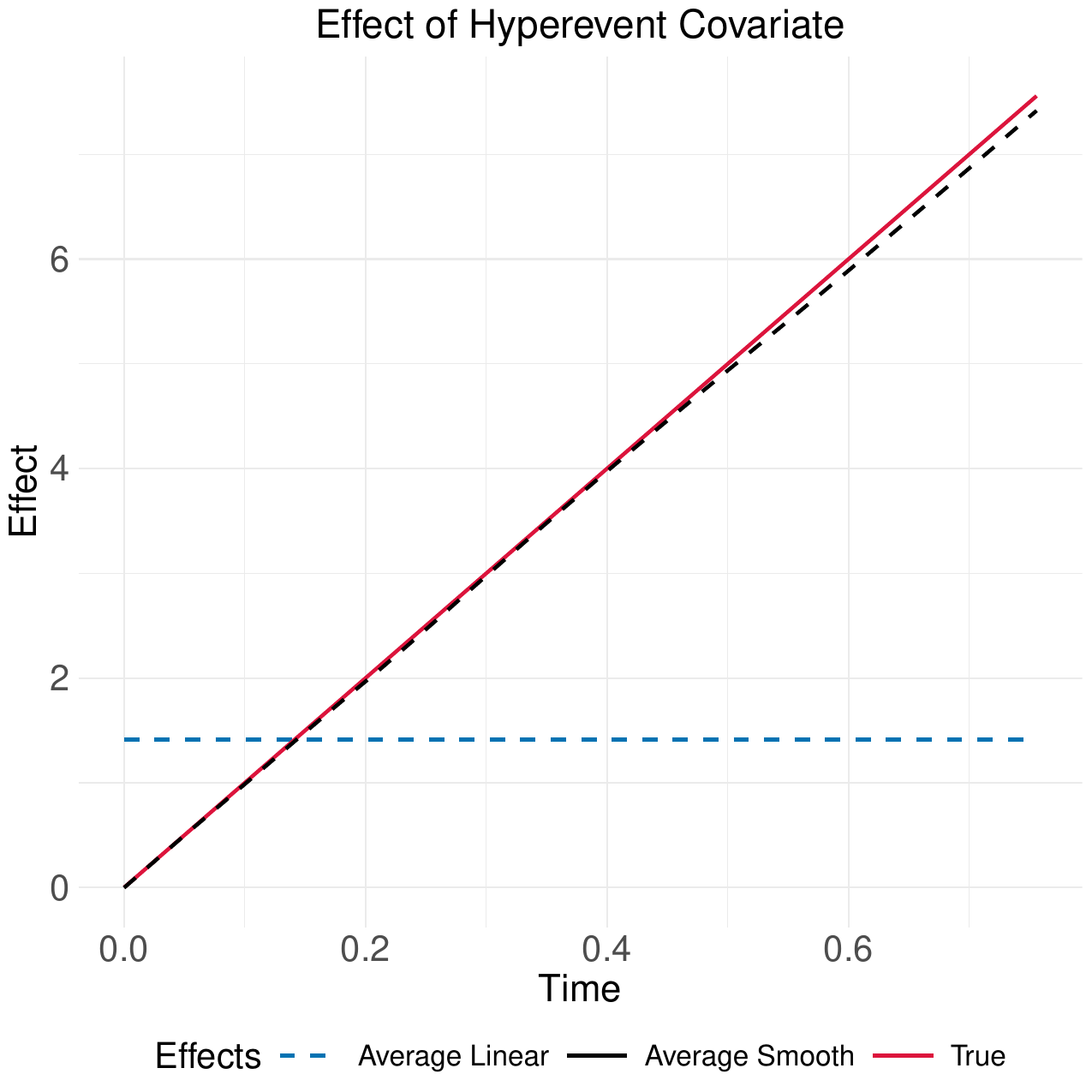} \\
    \end{minipage}
    \begin{minipage}{0.45\linewidth}
        \centering
        b) TVNLE Model \\
        \includegraphics[width=\linewidth]{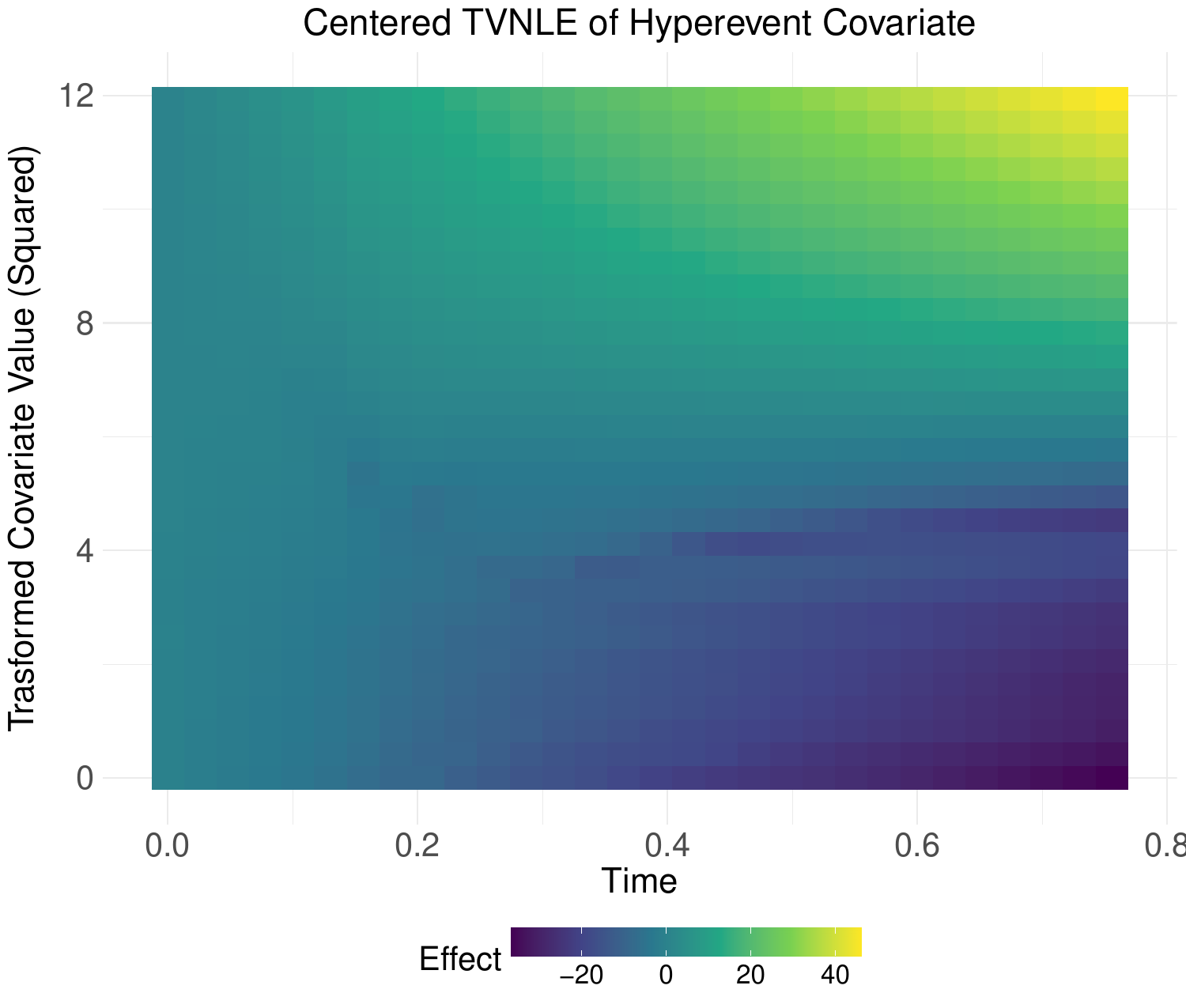} \\
    \end{minipage}
    \caption[Linear Models Fail to Capture Non-Linear Effects.]{\label{fig:simulation_study_time_varying}\textbf{Linear Models Fail to Capture Non-Linear Effects.}  
    Panels a) and b) display results based on synthetic data generated according to the model in Equation~\ref{eq_RG2-3}. Panels a) shows estimates from linear and time-varying models, while panel b) presents results from a model incorporating a joint time-varying and non-linear effect. 
    a) Estimates from multiple replications are combined using inverse variance weighting\index{inverse variance weighting}. For the linear model, we aggregate the estimated slopes to create a consensus line, called the \enquote{consensus linear effect} (blue dashed line). Unlike Figure~\ref{fig:simulation_study_linear}, the x-axis here shows time. As a result, the consensus linear effect appears as a horizontal line, since it does not change over time. In contrast, the true effect (red solid line) changes over time. Linear models cannot capture this time-dependent pattern. For time-varying effects, we first interpolate predicted effects across time points. Then, we aggregate them pointwise, weighting each by the inverse of its variance. This produces a \enquote{consensus time-varying effect} (black dashed line) that captures the average time trend well. Notably, unlike in Figure~\ref{fig:simulation_study_linear}, the sign of the effect here can be interpreted. Therefore, the consensus non-linear effect is not shifted. b)  For the TVNLE model, estimates are aggregated using inverse variance weighting. The resulting smooth surface is then centered such that, at each time point, the average effect across the covariate domain is equal to \(0\).}
\end{figure}

\subsection{Linear models fail to capture time-varying effects.}
{In this section we simulate data with various time-varying effects $\gamma_j(t)$. The aim is to show that the traditional REM model fails to capture this aspect, whereas our REM extension correctly captures its time-varying nature.}
We simulate hyperevent data from the true underlying model with time-varying effects, 
\begin{equation}\label{eq_RG2-3}
    \log\lambda(t, I) = \underbrace{-10 \cdot t}_{{\gamma_1(t)}} \cdot \log\left(\sqrt{\text{subrep}^{1}(t, I)}\right) \underbrace{+ 10 \cdot t}_{{\gamma_2(t)}} \cdot \log\left(\sqrt{\text{subrep}^{2}(t, I)}\right) + \underbrace{+ 10 \cdot t}_{{\gamma_3(t)}} \cdot \overline{x}(I)^2 - 0.1 \cdot |I|.
\end{equation}
{Here, \(\gamma_k(t)\) for \(k = 1, 2, 3\) are time-varying functions. By contrast, the true effect of event size is time-invariant and linear. More specifically, we chose a time-varying function that is linear in time, i.e.\ \(\gamma_1(t) = -10t, \gamma_2(t) = 10t, \gamma_3(t) = 10t\). In principle, it is possible to simulate more complex functions, such as the non-monotonic example in Figure \ref{fig:example_time_varying}. However, the goal here is to show that even with a simple time-varying function, a linear effect fails to capture the true trend. As in Equation \ref{eq_RG1}, the coefficients \(\gamma_1(t) = -10t\) and \(\gamma_2(t) = 10t\) were chosen to balance each other, preventing the rate from exploding as events accumulate.}

Figure~\ref{fig:simulation_study_time_varying} shows that the linear model fails to capture the time-varying effect of this variable. By contrast, the TVE model successfully identifies it. The TVNLE model also correctly identifies the time-varying -- but, for each time point, linear -- effect. As can be seen, for time equal to zero, the effect does not increase with the covariate (constant color on the vertical line at $t=0.0$), while the vertical color gradient becomes steeper as time increases.

\subsection{The potential of the joint time-varying non-linear estimation.}

In the previous analysis, we assumed knowledge of the true non-linear transformation of the covariate, applying it explicitly in the model. This is clearly unrealistic in practice. {In this third section, we therefore do not assume to know the true square transformation of $\bar{x}(I)$ nor its time-varying nature and instead allow the transformation $f$ to be learned from the data itself, permitting it to vary over both $\bar{x}(t)$ and time $t$, i.e., $f(\bar{x}(t), t)$.}

\begin{figure}
    \centering
    \begin{minipage}{0.75\textheight}
        \centering
        \begin{tabular}{cc}
          a) & b) \\
		  \includegraphics[width=0.4\linewidth]{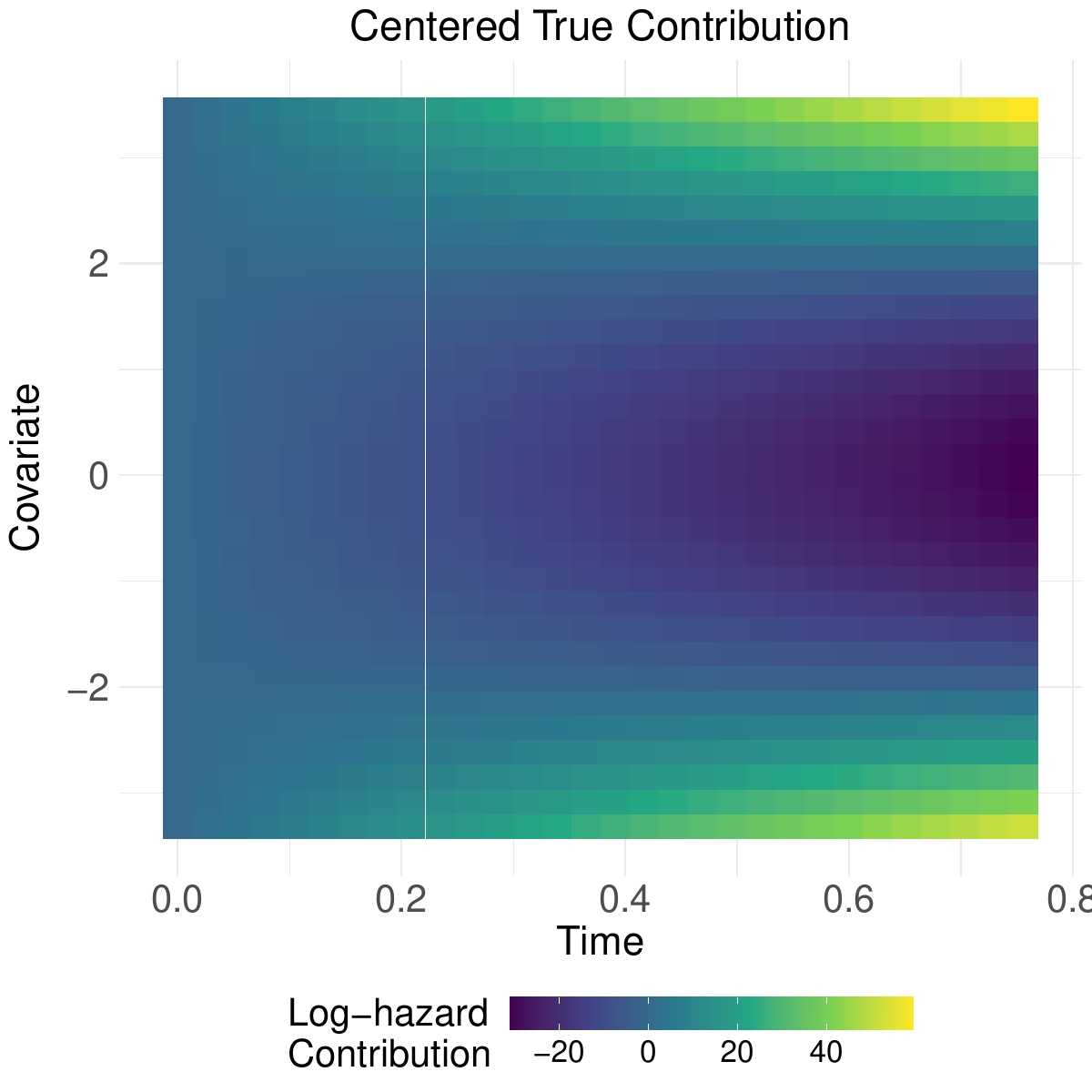} & 
		  \includegraphics[width=0.4\linewidth]{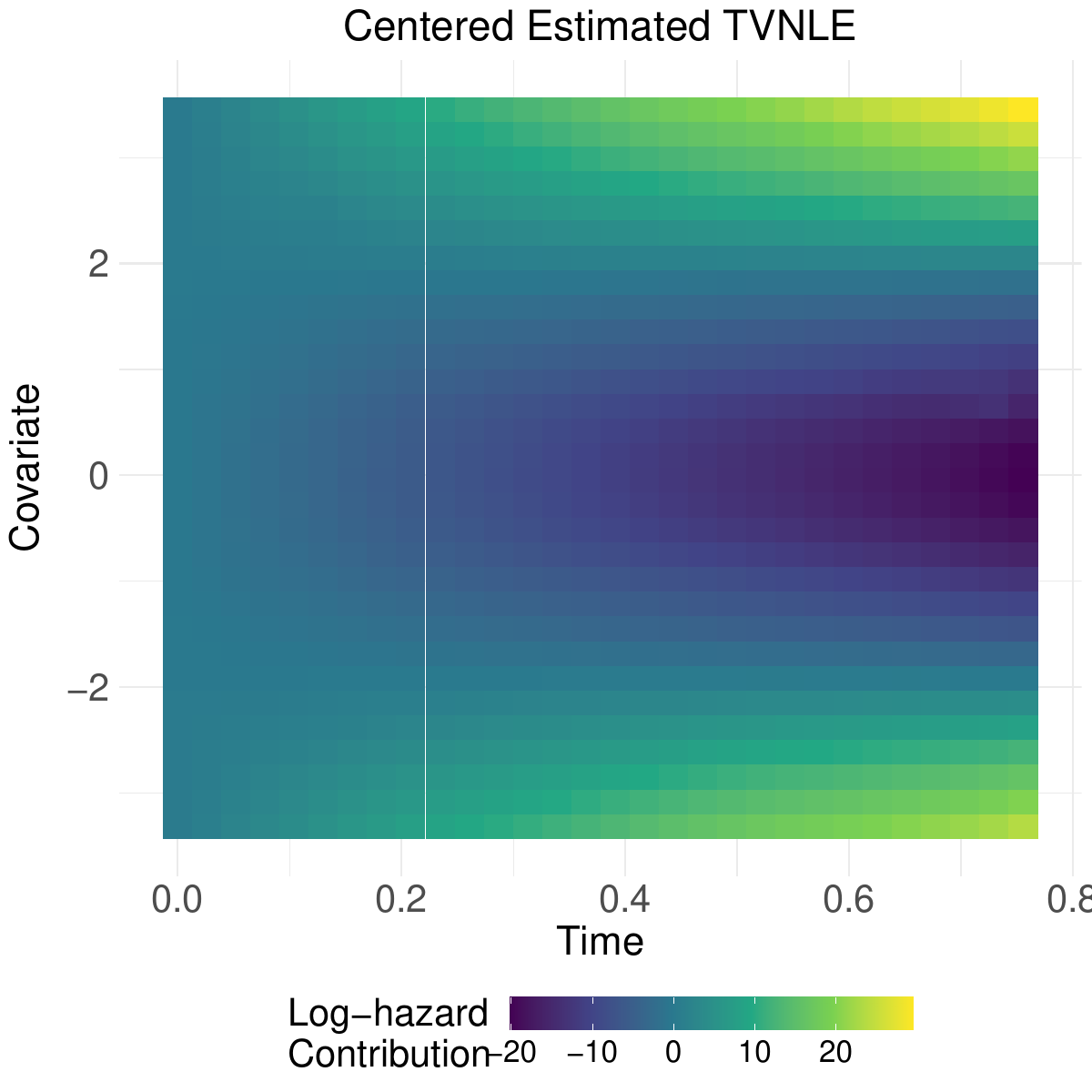}\\
          c) & d) \\
		  \includegraphics[width=0.4\linewidth]{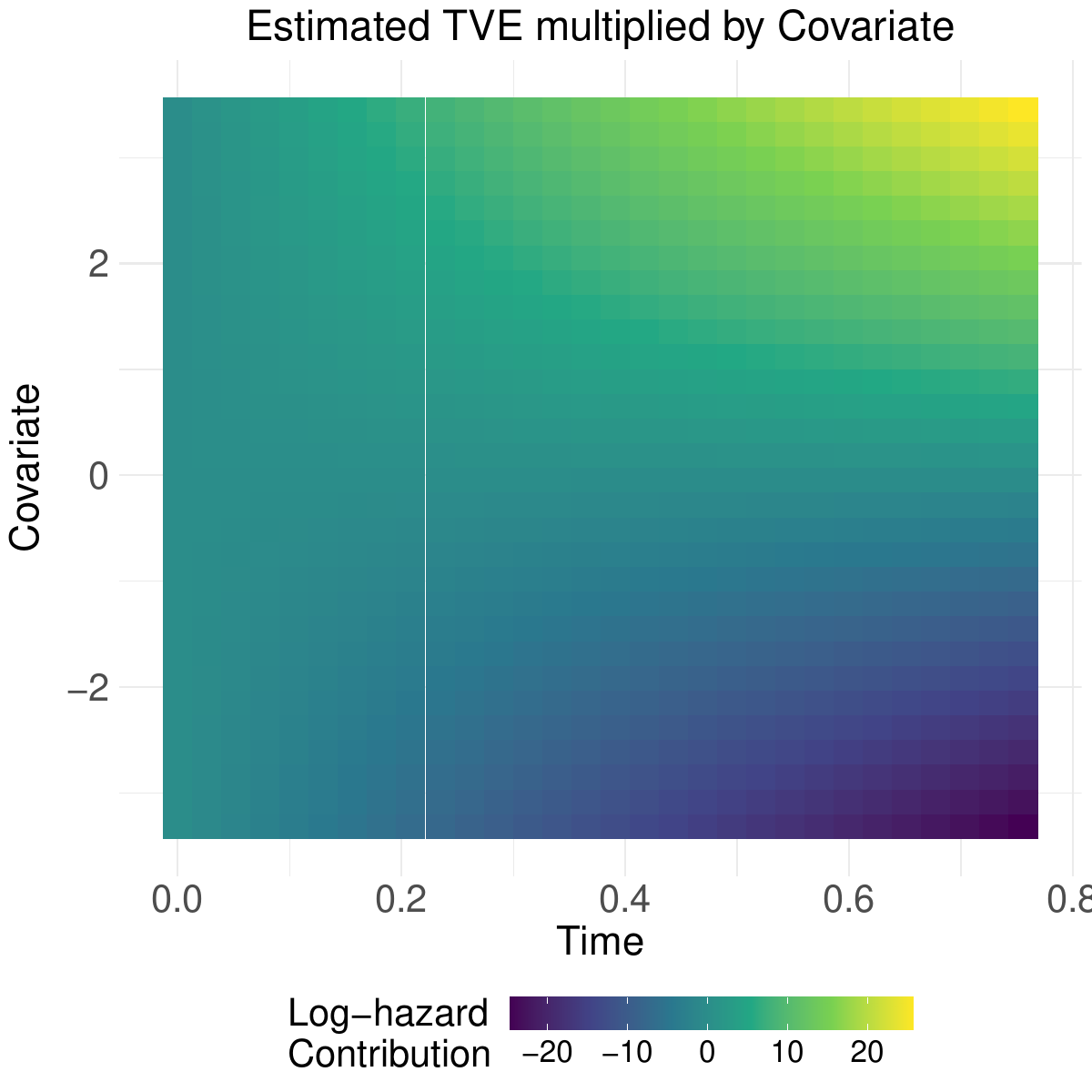} &
		  \includegraphics[width=0.4\linewidth]{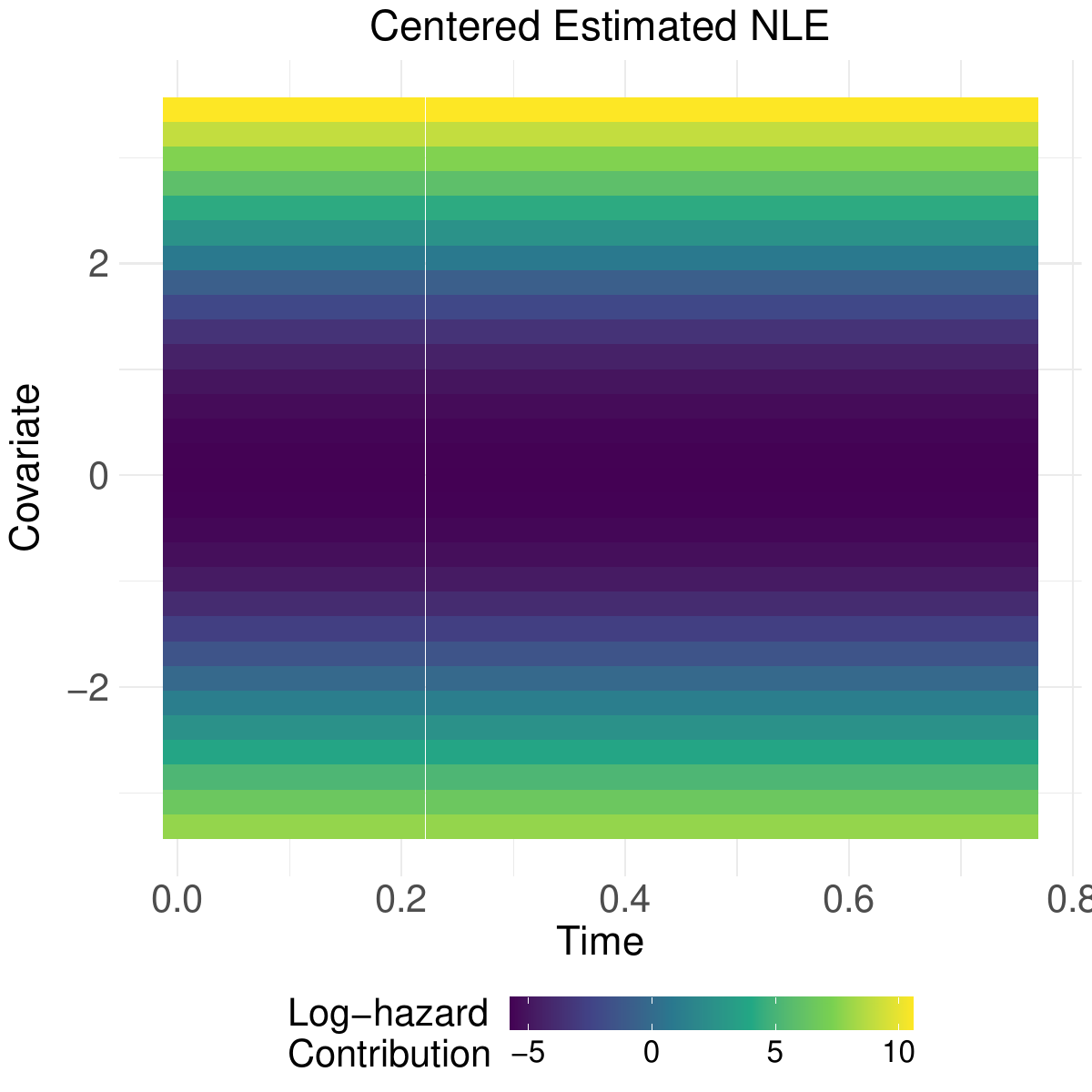}
	   \end{tabular}
    \end{minipage}
    \caption[The potential of joint time-varying non-linear estimation and the failure of time-varying and non-linear models.]{\label{fig:simulation_study_time_varying_non_linear}\textbf{The potential of joint time-varying non-linear estimation and the failure of time-varying and non-linear models.} The four plots display results from simulated data based on the model in Equation~\ref{eq_RG2-3}. The two upper plots, a) and b), show centered true effect and the estimated effect, respectively. The color shading represents the effect \(f^{\text{TVNLE}}(\overline{x}(I), t)\), with time \(t\) on the x-axis and the values of the hypercovariate \(\overline{x}(I)\) on the y-axis. The two lower plots, c) and d) show interpolated values of the estimated contributions to the log-rate function, obtained by fitting a model with only time-varying effects and non-linear effects, respectively. The contribution to the log-rate with TVE consists of the TVE itself multiplied by the value of the covariate at the corresponding time.}
\end{figure}

Figure~\ref{fig:simulation_study_time_varying_non_linear} a) shows the true centered effect, i.e. \(f^{\text{TVNLE}}\left(\overline{x}(I), t\right) = + 10 \ t \ \overline{x}(I)^2 \) and b) the estimated effect \(\hat{f}^{\text{TVNLE}}\left(\overline{x}(I), t\right) \). The color shading represents the effect, plotted with time \(t\) on the x-axis and the values of the hyperevent covariate \(\overline{x}(I)\) on the y-axis. We observe that the estimates successfully capture both the variation over time and the variation across covariate values. The image was generated by predicting the log-hazard contributions for 30 pairs of time and covariate values, uniformly sampled from the range of the synthetic data.

\subsection{Time-varying and non-linear models fail to capture simultaneous time-varying non-linear effects.}

{Using the same simulation following the model in \eqref{eq_RG2-3}, this section aims to show that joint time-varying and non-linear estimates are strictly necessary to uncover the underlying structure of the dynamics. Only time-varying or only non-linear effects fail dramatically, leading to wildly different conclusions.}

Figure~\ref{fig:simulation_study_time_varying_non_linear} c) and d) display interpolated values of the estimated contributions to the log-rate function, obtained by fitting a model with time-varying effects and non-linear effects only. It is important to note that when considering TVE, the log-hazard contribution corresponds to the time-varying effect multiplied by the value of the covariate at the corresponding time. We observe that the TVE contribution incorrectly identifies a decrease in the value of the contribution for low covariate values towards the end of the time window. In contrast, the NLE model fails to capture the temporal variation in the data.
    
    % EMPIRICAL APPLICATION
    \section{Empirical Application to Coauthorship Citation Networks} 
\label{sec_empirical_application}

\begin{figure}
\centering

\begin{adjustbox}{max width=\textwidth}

\begin{tikzpicture}[node distance=2.5cm]

\node[mainblock] (eventnet)
    {\textbf{1. Compute covariates for events and non-events using \texttt{eventnet}}};

\node[subblock, below=0.2cm of eventnet] (eventnetsub) {
Using the \texttt{eventnet} software,
\begin{enumerate}
    
    \item Extract event information from the raw data.
    
    \item Specify the components required by \texttt{eventnet} to define the risk set.
    
    \item Sample a non-event observation from the risk set.

    \item Compute the covariates for both events and non-events.
    
\end{enumerate}

A tutorial and example code for creating an \texttt{eventnet} configuration for publication event data is provided at the following \href{https://github.com/juergenlerner/eventnet/wiki/Coevolution-of-collaboration-and-references-to-prior-work-(tutorial)}
{Link}.

};

\node[mainblock, below=1.5cm of eventnetsub] (casecontrol)
    {\textbf{2. Create the case-control dataset}};

\node[subblock, below=0.5cm of casecontrol] (casecontrolsub) {

Merge the output of \texttt{eventnet} by event and non-event identifiers so that each row contains information for both the event and the non-event.

};

\node[mainblock, below=1.5cm of casecontrolsub] (transform)
    {\textbf{3. Transform covariates and event times}};

\node[subblock, below=0.5cm of transform] (transformsub) {
\begin{enumerate}
    \item Apply a monotonic transformation to each covariate $x$:

    \[
    \dot{x} = 1 - \exp\left(-\frac{x}{c}\right)
    \]

    Notes:
    \begin{itemize}
        \item Choose \(c\) so that the transformed covariate (evaluated for events) is as close as possible to a uniform distribution.
        \item Apply the same transformation to both events and non-events.
    \end{itemize}

    \item Transform event times $t$ using the empirical cumulative distribution function:

    \[
    \dot{t} = \mathrm{ecdf}_t(t)
    \]
\end{enumerate}
};

\node[mainblock, right=3cm of transform] (gam)
    {\textbf{4. Fit the model using \texttt{mgcv::gam()}}};

\node[subblock, below=0.5cm of gam] (gamsub) {
\begin{verbatim}
n = nrow(data)

X = cbind(data$transf_x_event,
               data$transf_x_non_event)

T = cbind(data$transf_time,
               data$transf_time)

unit = rep(1, n)
I = cbind(unit, -unit)

gam(y ~ -1 + te(T, X, by = I),
      family = "binomial")
\end{verbatim}
};

\node[mainblock, right=3cm of eventnet] (inference)
    {\textbf{5. Inference and diagnostics}. Interpret the obtained time-varying non-linear contributions. };

\node[subblock, below=0.5cm of inference] (inferencesub) {

    \textbf{Example: AAS}.  
    This effect exhibits a non-monotonic pattern both over time and across covariate values.  
    A detailed interpretation is provided in Figure~\ref{fig:example_tvnle}.
  
  \begin{center}
      \includegraphics[width=0.6\textwidth]{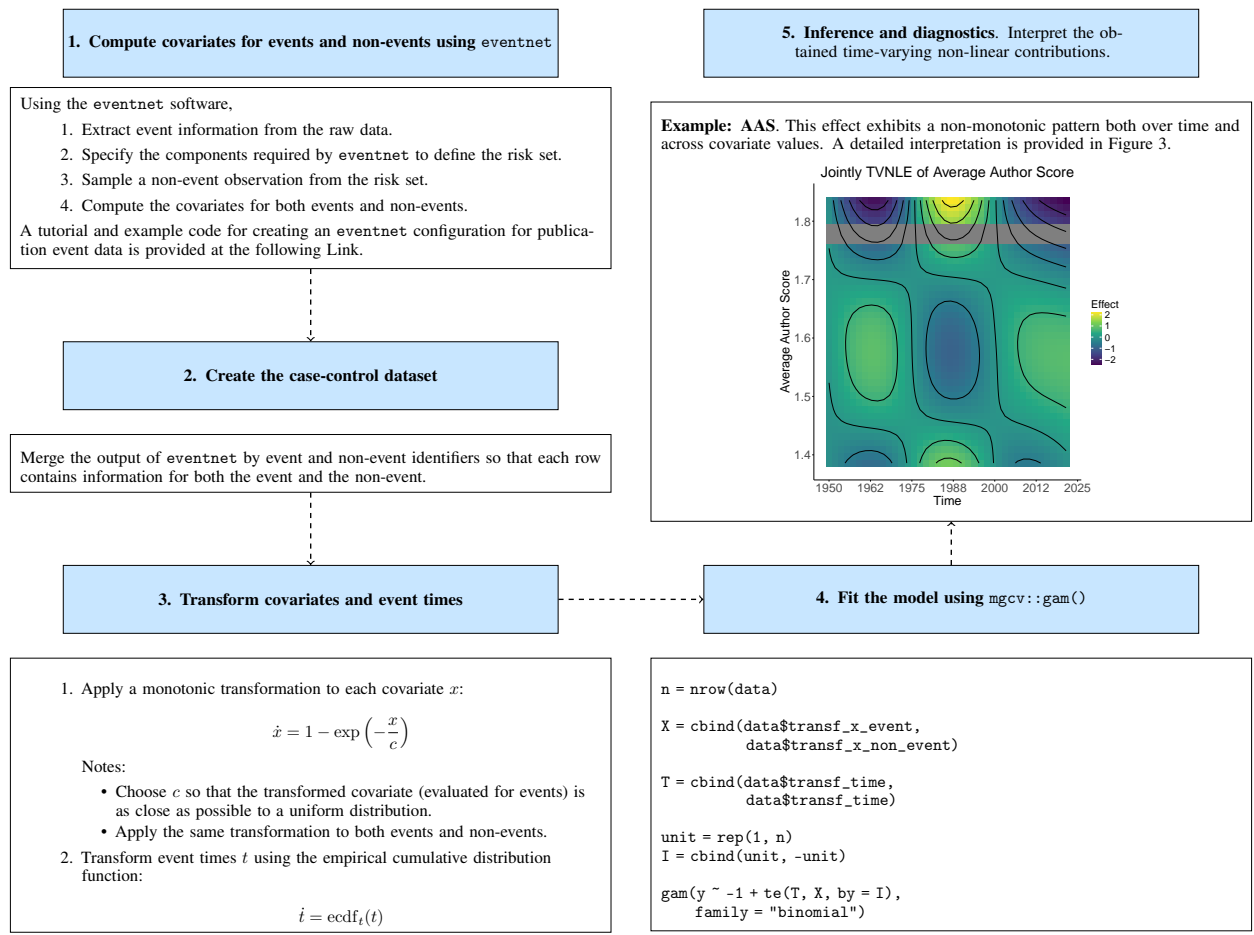}
  \end{center}
};

% === ARROWS ===
\draw[arrow] (eventnetsub) -- (casecontrol);
\draw[arrow] (casecontrolsub) -- (transform);
\draw[arrow] (transform) -- (gam);
\draw[arrow] (gam) -- (inferencesub);

\end{tikzpicture}

\end{adjustbox}

\caption{\label{fig_flow_chart} Flow chart summarizing the steps required to replicate the analysis.}
\end{figure}

We applied our RHEM extension to the DBLP-Citation-network V14 dataset, restricted to journal articles, derived from the Aminer Citation Network \citep{tang2008arnetminer}. 
In this context, a relational hyperevent is defined as the formal \textit{publication of a scientific work}\index{publication of a scientific work}, marking the moment when a group of authors presents their research alongside a set of citations listed in the reference section. Our analysis is specifically focused on journal publications, constructing an event list that includes only these types of works. Furthermore, among the cited works, we also restrict our selection to journal publications to maintain consistency within the dataset. 
Code to pre-process and filter the data is publicly available.\footnote{\url{https://github.com/juergenlerner/eventnet/tree/master/data/scientific_networks/aminer_2023}} 

The resulting event sequence comprises \(n=1,416,353\) publication events spanning from 1939 to 2023.
\begin{equation*}
	\{ j_m = (t_m, I_m, J_m), m = 1, \ldots, 1\ 416\ 353 \}
\end{equation*}
Here, $j_m$ represents the journal article published at time $t_m$ by a group of authors in \( I_m \subseteq V^I_{t_m} \) citing journal articles in \( J_m \subseteq V^J_{t_m} \). {The events in the empirical data can be represented exactly as in Figure \ref{fig:example_introduction}, namely as time-stamped directed hyperedges of a two-mode network. There are two types of nodes: authors and cited papers. When a new hyperedge occurs at time \(t_m\), a new publication \(j_m\) -- authored by a set of authors in \(I_m\) and citing a set of papers in \(J_m\) -- enters the system and can be cited by other papers at later time points. This two-mode network representation also makes it possible to examine more complex relationships, such as author-cites-paper, paper-cites-paper, and author-cites-author. These three relations are part of a broader set used to compute the covariates defined in Section \ref{subsec_drivers}. For this reason, labeling this structure simply as a bipartite network in which authors are connected to the papers they cite may be an oversimplification.  Specifically, the counting process \(N(t, \cdot, \cdot)\) tracks the cumulative number of papers produced by authors citing existing works  up to time \(t\). This approach allows us to examine how endogenous relational mechanisms -- particularly prior scientific collaborations or citation patterns among authors -- influence the publication rate of journal articles \(\lambda(t, \cdot, \cdot)\). The co-evolution of co-authoring and citation networks was first proposed by \citep{lerner2024relational}. In this paper, we relax their assumption of linear and time-homogeneous effects, showing that such flexibility is warranted for these data.}

{The empirical data under consideration spans the period from 1939 to 2023. Event times are recorded with yearly granularity, so many papers appear as occurring simultaneously. From a theoretical point of view, this does not imply that these events actually occurred at the same moment (indeed, it is likely not the case). However, because we lack information to determine their order, some adjustments are required. First, when building the case-control data set, the sampled non-hyperevent must differ not only from the corresponding hyperevent but also from all hyperevents occurring in the same year. Second, when computing endogenous covariates, publication events from the same year cannot be treated as previously occurred events. In practice, the \texttt{eventnet} software accounts for this by allowing the user to specify the variable that identifies simultaneous events, such as the yearly event time recorded in this data set.}

{Event times are also not uniformly distributed, which further motivates the inclusion of time-varying effects. Indeed, over 87\% of the publications occurred starting from 2000.}
By including in the model formulation an explicit reference to the time of the event, interacting with the covariate evaluated at the time of interest, our technique allows us to account for this temporal imbalance. 
Moreover, this empirical application carries significant social implications. Scientific networks provide insights into the factors motivating different authors to collaborate, revealing the complex dynamics that drive co-authorship and knowledge exchange. This presents a valuable opportunity to deepen our understanding of the evolving dynamics within complex social networks, particularly within the scientific domain. These networks consist of interconnected authors and papers, but they can also include inventors, awardees, investors, publications, grants, and patents, along with their intricate relationships. Such networks are continually expanding, embodying social phenomena like the exchange of funding, knowledge, and reputation. There is no inherent justification for assuming that this influence is linear or homogeneous over time, as collaboration patterns evolve in response to shifts in research interests, available resources, and community structures, as demonstrated in various approaches to analyzing co-authorship networks \citep{hoekman2010research, kwiek2021large}. {Other motivations for this study can be found in Section \ref{sec_motivation}. A flow chart summarizing the steps required to replicate the analysis is provided in Figure \ref{fig_flow_chart}.}

\subsection{Drivers of Scientific Collaboration and Impact}\label{subsec_drivers}

Using the case-control partial likelihood\index{likelihood!case-control partial likelihood} inference framework with GAMs, as explained in Section~\ref{sec_inference_procedures}, we sampled one possible but unobserved event from the risk set for each publication event. For each group of authors \( I \) and the corresponding group of cited papers \( J \) at time \( t \), we selected a pair \( (I^\ast, J^\ast) \in \binom{V^I_{t}}{|I|} \times \binom{V^J_{t}}{|J|} \), ensuring both groups have the same size in the hyper-event and the non-hyperevent. Hyperedge covariates were generated using the open-source software \texttt{eventnet} \cite{lerner2020reliability,lerner2023relational},\footnote{\url{https://github.com/juergenlerner/eventnet}} which allows sampling a specified number of non-event hyperedges related to each observed publication event, and calculating covariates for both observed events and sampled non-events. In our analysis, we sample one non-event (\(m=1\)).

In this empirical application, relational hyperevents represent \textit{citations}\index{citation}. Therefore, we replace the generic term \enquote{action} in Equation~\eqref{eq_action} with \enquote{citation}. We focus on a subset of the hyperedge covariates studied in \cite{lerner2024relational} and, following that work, incorporate a weight factor \(\omega\) to model the temporal decay of past events. Most of the statistics used can be defined by adapting and combining the function \enquote{citation}. Specifically, we introduce the following adaptations, based on the superscript referring to either authors or papers (\(\textup{aut}\) or \(\textup{pap}\)):

\begin{enumerate}
    \item \textit{Author-Paper citations}\index{Author-Paper citations}\label{item-acp}:
    \[
    \textup{cite}^{\textup{aut}-\textup{pap}}(t, I, J) = \sum_{t_m < t} \omega(t-t_m) \cdot \mathbbm{1}_{\{I \subseteq I_m \cap J \subseteq J_m\}} 
    \]
    \item \textit{Paper-paper citations}\index{Paper-paper citations}\label{item-pcp}::
    \[ 
    \textup{cite}^{\textup{pap}-\textup{pap}}(t,j,j') = \sum_{t_m<t} \omega(t-t_m) \cdot \mathbbm{1}_{\{j=j_m\wedge j'\in J_m\}}
    \]
    \item \textit{Author-Author citations}\index{Author-Author citations}\label{item-aca}::
    \[ 
    \textup{cite}^{\textup{aut}-\textup{aut}}(t,i,i') = \sum_{t_m<t} \omega(t-t_m) \cdot \mathbbm{1}_{\{i\in I_m \ \wedge \ \exists t_{m'} \ \colon \ j_{m'} \in J_m \ \wedge \ i'\in I_{m'}\}}
    \]
    \item \textit{Author citation popularity}\index{Author citation popularity}:
    \[ 
    \textup{cite\_pop}^{\textup{aut}}(t,i) = \sum_{t_m<t} \omega(t-t_m) \cdot \mathbbm{1}_{\{ \exists t_{m'} \ \colon \ j_{m'} \in J_m \ \wedge \ i\in I_{m'} \}}
    \]
    \item \textit{Authorship}\index{Authorship}:
    \[ 
    \textup{author}(t, i, j) = \sum_{t_m<t} \omega(t-t_m) \cdot  \mathbbm{1}_{\{ j = j_m \ \wedge \ i\in I_m \}}
    \]
    \item \textit{Out-degree}\index{Out-degree}:
    \[ 
    \textup{out\_degree}(j) = \sum_{t_m} \omega(t-t_m) \cdot  \mathbbm{1}_{\{ j = j_m \}} \cdot |J_m|,
    \]
\end{enumerate}
{where \(\omega(t-t_m) = \exp\Big\{ - (t-t_m) \frac{\log 2}{T_{\frac{1}{2}}} \Big\}\) and \(T_{\frac{1}{2}}>0\) is the half-life period \citep{lerner2013modeling}, set to three years in this empirical application, following \citet{lerner2024relational}}.

As we show that ``action'' defined in Equation \eqref{eq_action} serves as a basic block for subset-repetition ``subrep'' defined in Equation \eqref{eq_subrepetition}, all the previously listed adaptations play a similar role. Specifically, we focus our attention on the following hyperedge covariates involving authors' network, papers' network and their interconnection (see \cite{lerner2024relational} for a more detailed description, graphical illustration, and numerical examples).

\begin{enumerate}
     \item \textit{Prior papers}\index{prior papers}:
    \[
    \textup{prior\_papers}(t,I,J)=\textup{subrep}^{(1,0)}(t,I,J)
    \]
     \item \textit{Prior joint papers}\index{prior joint papers}:
    \[
    \textup{prior\_joint\_papers}(t,I,J)=\textup{subrep}^{(2,0)}(t,I,J)
    \]
     \item \textit{Paper citation popularity}\index{paper citation popularity}:
    \[
    \textup{paper\_citation\_popularity}(t,I,J)=\textup{subrep}^{(0,1)}(t,I,J)
    \]
     \item \textit{Paper pair co-citation}\index{paper pair co-citation}:
    \[
    \textup{paper\_pair\_cocitation}(t,IJ)=\textup{subrep}^{(0,2)}(t,I,J)
    \]
     \item \textit{Author citation repetition}\index{author citation repetition}:
    \[
    \textup{author\_citation\_repetition}(t,I,J)=\textup{subrep}^{(1,1)}(t,I,J)
    \]
     \item \textit{Cite paper and its references}\index{cite paper and its references}:
    \[
    \textup{cite\_paper\_and\_its\_references}(t,I,J)=\sum_{\{j,j'\}\in {\binom{J}{2}}} \dfrac{\textup{cite}^{\textup{pap}-\textup{pap}}(t,j,j')+\textup{cite}^{\textup{pap}-\textup{pap}}(t,j',j)}{\binom{|J|}{2}}
    \]
     \item \textit{Difference in prior papers}\index{difference in prior papers}:
    \[
    \textup{difference\_in\_prior\_papers}(t,I,J)=\sum_{\{i,i'\}\in{\binom{I}{2}}} \frac{|\textup{cite}^{\textup{aut}-\textup{pap}}(t,\{i\},\emptyset)-\textup{cite}^{\textup{aut}-\textup{pap}}(t,\{i'\},\emptyset)|}{{\binom{|I|}{2}}}
    \]
     \item \textit{Author citation popularity}\index{author citation popularity}:
    \[
    \textup{author\_citation\_popularity}(t,I,J)=\sum_{i\in I}\frac{\textup{cite\_pop}^{\textup{aut}}(t,i)}{|I|} 
    \]
     \item \textit{Difference in author citation popularity}\index{difference in author citation popularity}:
    \[
    \textup{difference\_in\_author\_citation\_popularity}(t,I,J)=\sum_{\{i,i'\}\in{\binom{I}{2}}}
    \dfrac{|\textup{cite\_pop}^{\textup{aut}}(t,i)-\textup{cite\_pop}^{\textup{aut}}(t,i')|}{{\binom{|I|}{2}}}
    \]
     \item \textit{Collaborate with citing author}\index{collaborate with citing author}:
    \[
    \textup{collaborate\_with\_citing\_author}(t,I,J)=\sum_{\{i,i'\}\in {\binom{I}{2}}}
\dfrac{\textup{cite}^{\textup{aut}-\textup{aut}}(t,i,i')+\textup{cite}^{\textup{aut}-\textup{aut}}(t,i',i)}{\binom{|I|}{2}}
    \]
     \item \textit{Author self citation}\index{author self citation}:
    \[
    \textup{author\_self\_citation}(t,I,J)=\sum_{i\in I,\;j\in J} \dfrac{\textup{author}(i,j)}{|I|\cdot |J|}
    \]
     \item \textit{Paper outdegree popularity}\index{paper outdegree popularity}:
    \[
    \textup{paper\_outdegree\_popularity}(t,I,J)=\sum_{j\in J} \frac{\textup{out\_degree}(j)}{|J|}
    \]
\end{enumerate}

{An implicit assumption in fitting any regression model using maximum likelihood estimation is that the covariate processes have finite second moments. Practically speaking, this means that for improved model fitting stability, the covariates should not have major outliers. \cite{lerner2024relational} employ a square-root transformation of the non-negative covariates. As this only provides partial stabilization, we modified the network statistics using a different approach. We map each of the covariates $x^{(k)}$ and the time variable $t$ from their original scale into the interval $[0,1]$ in a way that distributes them as uniformly as possible.} 

{Specifically, we transform the event time variable using its \textit{empirical cumulative distribution function}\index{empirical cumulative distribution function} (ECDF). 
\begin{equation*}
    \dot{t} = \text{ecdf}_t(t)
\end{equation*}
where \( \text{ecdf}_t \) is the ECDF of the event times. This transforms the event times exactly evenly spaced into $[0,1]$.}

{For each covariate, \(x^{(k)}\), we propose a monotone increasing transformation into \([0,1]\):
\begin{equation}
\label{eq:covariate_transformation}
    \dot{x}^{(k)} = 1 - \exp\left(-\frac{x^{(k)}}{c^{(k)}}\right),
\end{equation}
where \(x^{(k)}\) denotes the value of the \(k\)-th covariate. The parameter \(c^{(k)}\) is chosen to minimize the Kolmogorov–Smirnov (KS) test statistic for uniformity of the covariate values evaluated at the events. Specifically, by varying \(c^{(k)}\), the empirical cumulative distribution function of the covariate changes and can be compared to a uniform distribution using the KS test, which evaluates the supremum of the difference between the two distributions. The KS statistic ranges from \(0\) to \(1\), and our goal is to minimize it. Once the optimal \(c^{(k)}\) is determined, the transformation in \eqref{eq:covariate_transformation} is applied to both event and non-event covariate values.}

{Although these transformations may seem to complicate interpretation, the values can be mapped back to the original scale, allowing the results to be interpreted in terms of the original covariates, while maintaining improved numerical stability in fitting the time-varying non-linear functions.}

\begin{table}[tb]
\centering
\begin{tabular}{rrr}
  \toprule
Excluded Covariate & AIC Difference & LogLik Difference\\ 
  \midrule
Difference in prior papers & -3886.01 & 1955.02 \\ 
  Author citation popularity & -2858.46 & 1435.35 \\ 
  Difference in author citation popularity & -408.09 & 215.52 \\ 
  Paper outdegree popularity & -433.07 & 232.89 \\ 
  Prior papers & -19288.83 & 9651.47 \\ 
  Prior joint papers (Linear) & -3210100.35 & 1605046.84 \\ 
  Collaborate with citing author & -341.42 & 176.77\\ 
  Paper citation popularity & -34990.31 & 17508.85 \\ 
  Paper pair cocitation & -15622.95 & 7809.36\\ 
  Author citation repetition & -6249.01 & 3134.45 \\ 
  Cite paper and its references & -19968.94 & 10001.37 \\ 
  Author self citation & -33584.8 & 16789.89\\ 
   \bottomrule
\end{tabular}
\caption{\textbf{Contribution of individual network attributes}. For each row in the table, we report the difference in AIC and log-likelihood (LogLik) between the full model (including all 12 covariates) and a reduced model that excludes the covariate listed in that row.}
\label{tab:AIC_contribution}
\end{table}

\subsection{Model fitting procedure}
Following the modeling framework outlined in Section \ref{sec_dynamic_hypernetwork_modeling}, we first assume a time-varying non-linear effect for each of the 12 covariates discussed in Section \ref{subsec_drivers}, contained in vector \(\bm{x}(t, I, J)\), and transformed according to Equation~\eqref{eq:covariate_transformation}. The contributions of these covariates to the log-hazard follows:
\begin{equation*}
	f\left(\dot{\bm{x}}(t, I, J), \dot{t}\right) = \sum_{k = 1}^{p=12} \sum_{q = 1}^{Q} \left( \sum_{l = 1}^L \alpha^{(k)}_{ql} a^{(k)}_{l}(\dot{t}) \right) b^{(k)}_{q}\left(\dot{x}^{(k)}(t, I, J)\right)
\end{equation*}
In this expression, \( f\left(\dot{\bm{x}}(t, I, J), \dot{t}\right) \) represents the contribution function\index{contribution function} expressing an additive joint time-varying non-linear effect of transformed covariates $\dot{\bm{x}}(t, I, J)$ on the log-hazard function. Each effect, in its trend, and not its sign, can be interpreted as the effect of the original covariate $x_k(t, I, J)$ on the log-hazard function. 

Fitting the complete model, we observed extreme values for the effect of the covariate Prior joint papers. On closer inspection, only a very small number of non-events took a value different from zero (18 out of 1,416,353). This indicates that, for a randomly selected potential event, it is rare to find a pair of actors with previous co-authorship. In contrast, events often show nonzero values for this covariate. To avoid unstable non-linear estimates, we chose to include a linear effect for this covariate.

To perform model selection, we explore only a subset of the nested models. Namely, for each of the 12 presented covariates, we fitted a model including all the other covariates. We then assessed the effect of each covariate by comparing the change in Akaike Information Criterion (AIC) and log-likelihood between the full and partial models. These differences can be evaluated in Table \ref{tab:AIC_contribution}. All twelve effects lead to an improvement in the AIC. Therefore, our final model contains all the mentioned covariates. To facilitate comparison with the baseline model assuming linear effects that are homogeneous over time, we fit standard RHEM \citet{lerner2024relational} specified with the twelve effects; see Table~\ref{tab:summary_baseline}.

\begin{table}[b]
\centering
\begin{tabular}{rrrrr}
  \toprule
 & Estimate & Std. Error & z value & Pr($>$$|$z$|$) \\ 
  \midrule
Difference in prior papers & 0.85 & 0.03 & 30.84 & <0.001 \\ 
  Author citation popularity & -1.14 & 0.03 & -39.33 & <0.001 \\ 
  Difference in author citation popularity & -0.70 & 0.03 & -23.09 & <0.001 \\ 
  Paper outdegree popularity & 0.32 & 0.01 & 24.49 & <0.001 \\ 
  Prior papers & -0.42 & 0.03 & -15.76 & <0.001 \\ 
  Prior joint papers & 440.00 & 21.12 & 20.83 & <0.001 \\ 
  Collaborate with citing author & 1.66 & 0.02 & 77.51 & <0.001 \\ 
  Paper citation popularity & 4.55 & 0.03 & 162.28 & <0.001 \\ 
  Paper pair cocitation & 40.58 & 1.37 & 29.56 & <0.001 \\ 
  Author citation repetition & 38.74 & 1.63 & 23.80 & <0.001 \\ 
  Cite paper and its references & 31.07 & 1.39 & 22.36 & <0.001 \\ 
  Author self citation & 47.56 & 1.92 & 24.82 & <0.001 \\ 
   \bottomrule
\end{tabular}
\caption{Summary of the baseline model. The baseline model include a linear effect for each of the 12 covariates discussed in Section \ref{subsec_drivers}, contained in vector \(\bm{x}(t, I, J)\), and transformed according to Equation~\eqref{eq:covariate_transformation}. The contribution function, in this case, is \(f^{\text{LE}}\left(\dot{\bm{x}}(t, I, J), \dot{t}\right) = \bm{\theta}^\top \dot{\bm{x}}(t, I, J)\).} 
\label{tab:summary_baseline}
\end{table}

\subsection{Evolving Dynamics of Scientific Collaboration and Impact}\label{subsec_resultss}

In the Supplementary Materials (Section B), we provide all figures illustrating the TVNL effects for the covariates included in the final model formulation. In this section, we focus on the interpretation of some of these effects showing a specific behavior: either non-monotone, time-varying, non-monotone and time-varying, or (approximately) monotone and homogeneous over time. We emphasize monotone vs.\ non-monotone, rather than linear vs.\ non-linear, since any linear approximation of a truly non-monotone effect would actually lead to a seriously invalid interpretation: any linear effect assumes that a covariate is either increasing or decreasing the event rate throughout its entire range -- while a non-monotone effect reveals that there are intervals in which the covariate increases the rate and other intervals in which it has a decreasing effect. One might argue that the practical implications of a non-linear, but strictly monotone effect are limited in many typical applications of relational event modeling, as analysts would still capture the correct direction of the effect.

It is important to note that the plots in the following figures should not be interpreted in terms of their sign. More precisely, the function $f^{\text{TVNLE}}\left(x(t, I, J), t\right)$ in Equation~(\ref{eq_tvnle}) is only identifiable up to the addition of a function $f_0(t)$ that may depend on time (but not on the covariate value $x(t,I,J)$). This is because any such function $f_0(t)$ would cancel out in each of the terms of the partial likelihood (\ref{eq_partial_likelihood}). Thus, using $f^{\text{TVNLE}}\left(x(t, I, J), t\right)+f_0(t)$ instead of $f^{\text{TVNLE}}\left(x(t, I, J), t\right)$ defines the same model. We have therefore centered the values displayed in the two-dimensional heatmap plots around zero for each time point. 

\begin{figure}[t]
    \centering
    \begin{minipage}{0.45\linewidth}
        \centering
        a) \\
        \includegraphics[width=\linewidth]{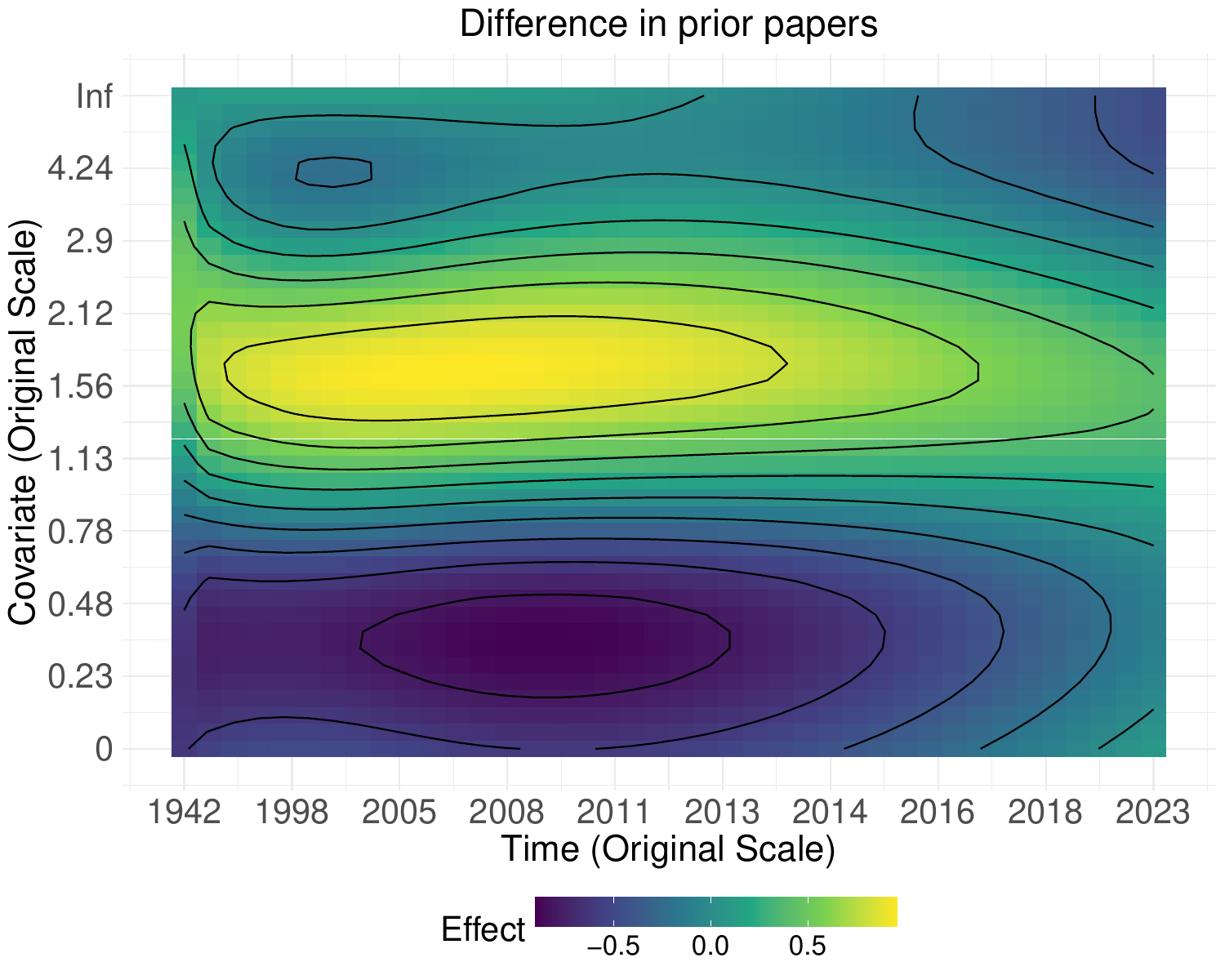}
    \end{minipage}
       \begin{minipage}{0.45\linewidth}
        \centering
        b) \\
        \includegraphics[width=\linewidth]{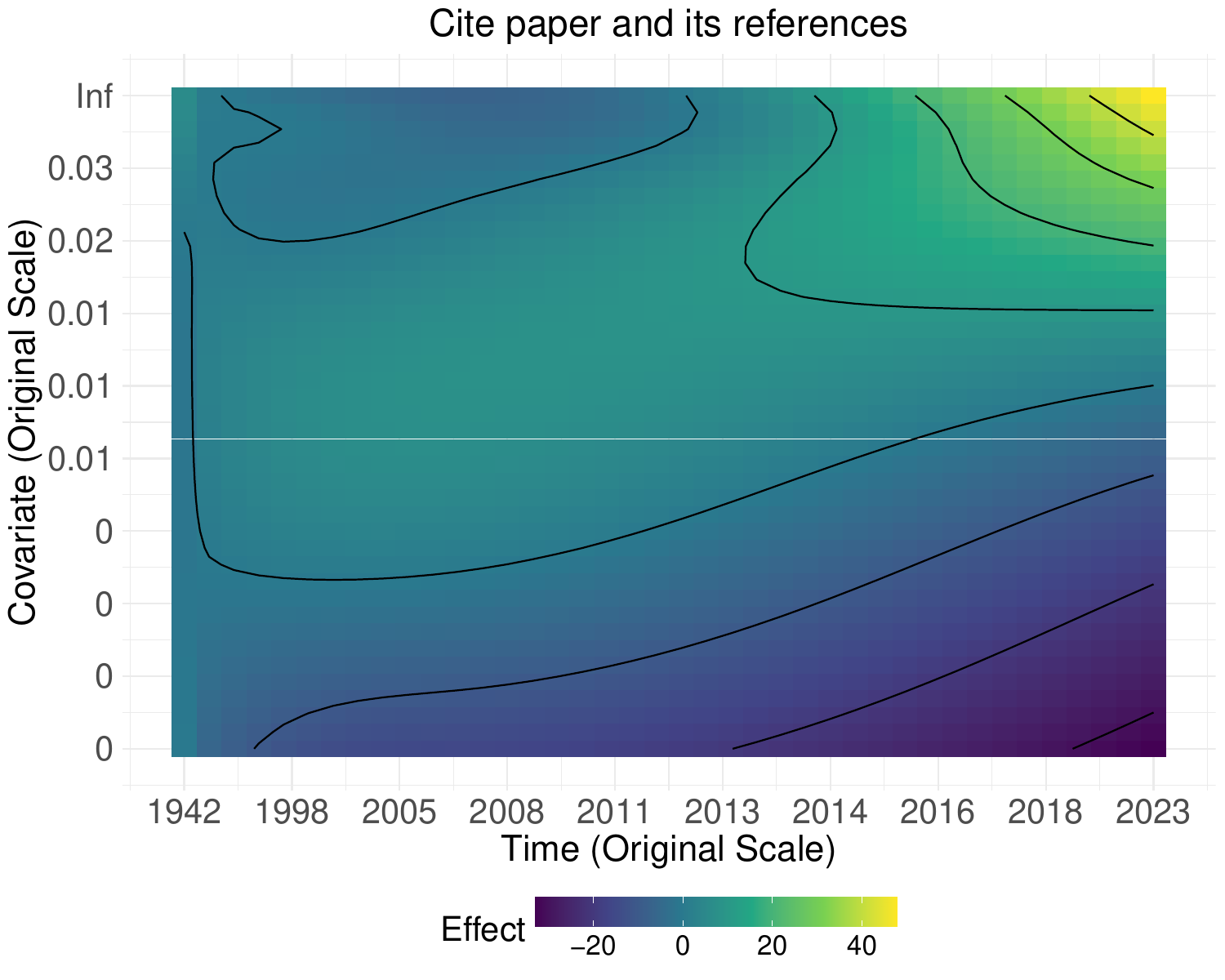}  
   \end{minipage}
    \caption[Estimated TVLE of Difference in prior papers and Cite paper and its references.]{\label{fig:empirical_effects1}\textbf{Estimated TVLE of Difference in prior papers and Cite paper and its references.} 
    a) Estimated TVNLE (log-hazard contribution) of Difference in prior papers. The plot should be interpreted in terms of trend, not sign. The estimate suggests a non-monotonic and thus non-linear effect across the range of the covariate. Although the effect appears nearly constant over time (except in recent years), for each fixed point in time, the pattern shows an initial decrease, followed by an increase, and then another decrease along the covariate range.
    b) Estimated TVNLE (log-hazard contribution) of Cite paper and its References. The effect is estimated as almost linear and very strong toward the end of the observation period. Before 2014, the covariate shows a non-monotonic pattern—first increasing and then decreasing with the covariate value. This illustrates a clear example of a time-varying effect.}
\end{figure}

\paragraph{Prior joint papers.} {We first consider the interpretation of the covariate \emph{prior joint papers}, which, as mentioned above, was included with a linear effect to avoid unstable non-linear estimates. The estimated fixed time-inhomogeneous coefficient for this variable, when the others are modeled with TVNLE, is 421.93 (s.e. 26.99). It is important to emphasize that this coefficient refers to the transformed covariate, whose values are mapped to the \([0,1]\) interval. Nonetheless, the positive effect remains interpretable as such, given the monotonic nature of the transformation and the fact that the original covariate takes only positive values.} Table~\ref{tab:AIC_contribution} highlights the importance of retaining this covariate in the model formulation. 

\paragraph{Difference in prior papers.} 
This covariate on a hyperedge $(I,J)$ at time $t$ is the average of the absolute pairwise differences in the number of previously published papers, taken over all pairs of authors in $I$. The baseline effect in Table~\ref{tab:summary_baseline} is positive, suggesting that teams of authors are typically composed of authors with large differences in the number of papers they have previously published, for example a PhD student together with her supervisor. The TVNLE plot in Figure~\ref{fig:empirical_effects1}a suggests a clearly non-monotone (hence non-linear) effect. If we fix a point on the horizontal axis (time) and go from bottom (covariate at its minimum value) to top (covariate at its maximal value), we find an increase in the event rate, up to about $x=1.5$, but then a decrease in the event rate. This means that the linear pattern (more difference in the number of prior papers implies a higher event rate) holds only for the lower values but gets reversed for higher values. The non-monotone pattern is qualitatively similar over time -- although the maximum is less pronounced at the end of the observation period and sharper at the beginning.

{The positive baseline effect reported in Table~\ref{tab:summary_baseline}, assuming a linear and homogeneous over time, suggests that scientists are more likely to coauthor a paper when potential team members differ more in their number of prior publications. In contrast, the non-linear effect in Figure~\ref{fig:empirical_effects1} a) indicates that there is an ``optimal'' level of variation in prior publications that maximizes this covariate’s contribution to the log-rate of coauthoring a new paper; beyond this point, the coauthoring rate declines -- here, ``optimal'' refers only to the co-authoring rate, not to paper quality.}

\paragraph{Cite paper and its references.} This covariate on a hyperedge $(I,J)$ at time $t$ is the average number of prior citations among all pairs of papers in $J$. The baseline effect in Table~\ref{tab:summary_baseline} is strongly positive, suggesting a tendency to partially copy the reference list of cited papers.
However, the TVNLE plot in Fig.~\ref{fig:empirical_effects1}b shows that this very strong effect mostly holds around the end of the observation period, while earlier the effect is much weaker. Prior to 2011 the covariate even has a slightly non-monotone effect that has an increasing effect on the event rate for low values of the covariate and a decreasing effect for high values. The tendency to cite a paper together with some of its references is thus an example of a time-varying effect. 

The finding that the tendency to cite a paper together with some of its references becomes very strong only at the end of the observation period could be of substantive interest as it gives support to the interpretation that increasing use of publication databases and paper-search technology might strengthen this effect \citep{lerner2024relational}. Since it becomes increasingly simple to follow paper citations forward and backward, and to retrieve the respective papers, the tendency to cite a paper and some of its references, or some of the papers previously citing it, may become stronger over time.

\begin{figure}[t]
    \centering
    \begin{minipage}{0.45\linewidth}
        \centering
        a) \\
        \includegraphics[width=\linewidth]{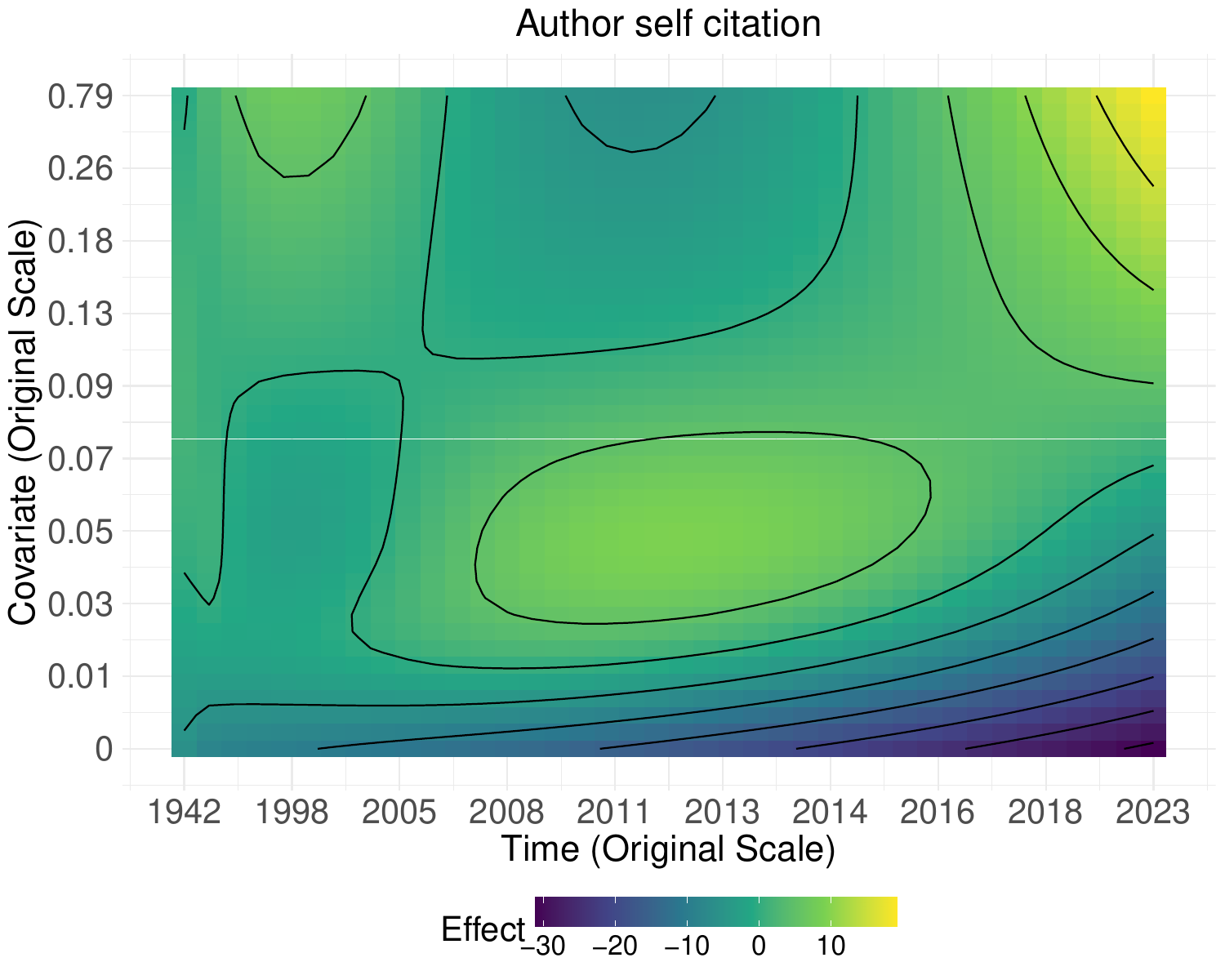}
    \end{minipage}
    \begin{minipage}{0.45\linewidth}
        \centering
        b) \\
        \includegraphics[width=\linewidth]{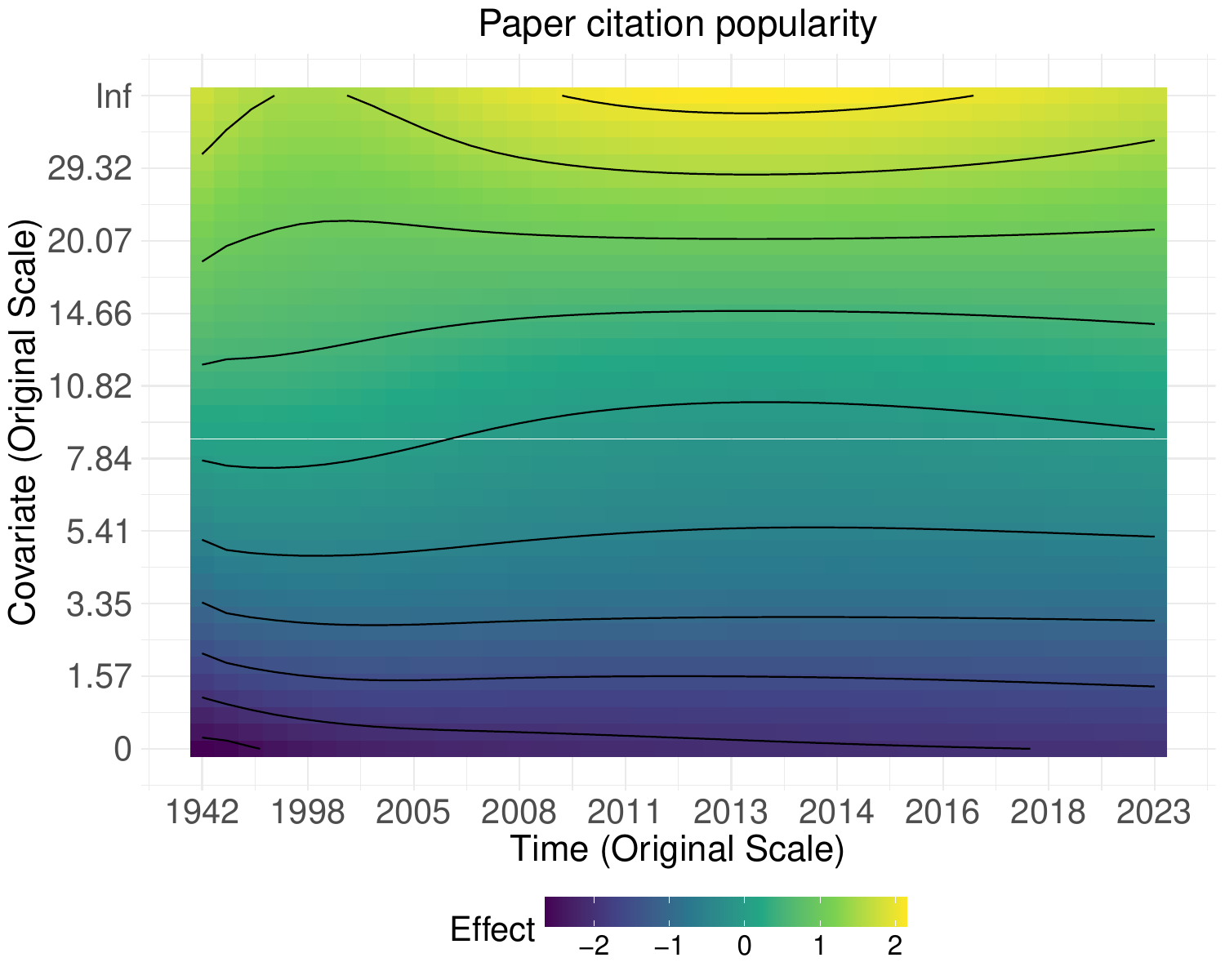}
    \end{minipage}
    \caption[Estimated TVLE of Author Self-Citation and Paper Citation Popularity.]{\label{fig:empirical_effects2}\textbf{Estimated TVLE of Author Self-Citation and Paper Citation Popularity.}  
    a) Estimated TVNLE (log-hazard contribution) of Author self-citation. The plot suggests a monotone effect at the beginning and end of the observation period. However, between 2008 and 2016, the effect is non-monotonic: it first increases with the covariate value, then decreases. In this case, using a TVNLE is essential, as the effect changes qualitatively over time and varies with the covariate.
    b) Estimated TVNLE (log-hazard contribution) of Paper Citation Popularity. The plot indicates a monotone effect that is approximately constant over time. For larger covariate values, the effect slightly varies across the time range, suggesting some mild time-dependence, but still being monotonically increasing.}
\end{figure}

\paragraph{Author self-citation.} 
This covariate on a hyperedge $(I,J)$ at time $t$ is the fraction of pairs $(i,j)\in I\times J$, such that $i$ is among the authors of $j$. The baseline effect in Table~\ref{tab:summary_baseline} is strongly positive, suggesting that authors have a tendency to cite their own prior work. The TVNLE plot in Fig.~\ref{fig:empirical_effects2}a suggests a more complex pattern where an approximately monotone increase can be found only at the beginning and end of the observation period, while for time in the interval $[2008,2016]$ the effect is clearly non-monotone, first increasing for low values of the covariate and then decreasing for high values. This covariate is thus an example of an empirical effect that is jointly time-varying and non-monotone.

{The non-monotone effect in the time interval $[2008,2016]$ suggests that during this time period possible reference lists with approximately 5\% to 7\% of self-citations have the highest probability to be the reference lists of actual publication events, rather than sampled reference lists of non-events. In contrast publications with more than 7\% self-citations become less likely during the time interval $[2008,2016]$. After 2016 the pattern becomes monotonically increasing, suggesting that in this period reference lists with higher ratios of self-citations are becoming increasingly more likely.}

\paragraph{Paper citation popularity.}
We have also found effects that are approximately monotone and nearly constant over time. The covariate Paper citation popularity on a hyperedge $(I,J)$ at time $t$ is the average number of prior citations received by the papers in $J$. The baseline effect in Table~\ref{tab:summary_baseline} is positive, revealing the well-known ``preferential attachment'' or ``rich-get-richer'' effect in which papers that have been cited more often in the past are cited at a higher rate in the future. The TVNLE plot in Fig.~\ref{fig:empirical_effects2}b qualitatively confirms this by showing a monotone effect of the covariate throughout the observation period (although the increase is somewhat steeper in the middle and flatter for the earlier and later time points). {The number of prior citations seems to monotonically increase a paper's likelihood to be cited in the future and that this pattern does hardly change over time.}

In summary, our empirical analysis suggests that in {relational hyperevent networks} we can find effects that are approximately monotone and constant over time (paper citation popularity), time-varying but almost monotone (cite paper and its references), non-monotone but mostly constant over time (difference in prior papers), and jointly time-varying and non-monotone (self-citation).

    % CONCLUSIONS
    \section{Discussion}
\label{sec_conclusions}

Whereas this paper focused on relational hyperevent models, also in the usual REM context, when dealing with dyadic relational events, one could consider incorporating jointly time-varying non-linear effects. Modeling and inference guidelines would remain largely the same. 

From a technical perspective, we have shown how tensor product smooths can be used to model time-varying non-linear effects dynamics within the framework of relational hyperevent models. We applied this approach to study the dynamics of citation and impact. Particularly in the context of the time-sensitive relational event process, this is a valuable addition in the network modeling toolbox. 

Jointly time-varying and non-linear effects are a flexible tool for discovering patterns that not only may be non-linear but also non-monotonic. 
Indeed, when assuming linearity of the effect, we implicitly assume that either the effect is always increasing or always decreasing. As shown in the empirical application presented in this paper, there are situations where the effect of a covariate reaches a maximum or minimum and then reverses its trend. For example, certain factors may encourage the occurrence of events up to a certain threshold, but beyond that point, they may become counterproductive. 
A concrete case of non-monotonicity is author self-citation {between 2008 and 2016}: citing a certain share of one’s own work seems natural and in fact may be nearly unavoidable, as one's own prior work may be among the most related to a new publication. However, excessive self-citation is perceived negatively by the research community and does not integrate the new work to that of others. 

{Certain patterns characterizing scientific collaboration and citations are evolving over time. In particular, several mechanisms exhibit changes in their behaviour towards the end of the observational window, such as self-citations and the tendency to cite chunks of the reference lists of cited works.
Other mechanisms, such as collaboration with a citing author and author-citation repetition, also show changes in recent years, either in trend or intensity, compared to earlier periods. Another notable example, described in the supplementary materials, concerns the impact of an author's prior publications. In the most recent period, the effect appears monotonic: the greater the number of prior publications, the more likely the author is to participate in a new publication. This contrasts with earlier periods, when teams with, on average, close-to-zero prior papers had a higher chance of publishing than those with a small but nonzero publication history. After reaching a minimum, the pattern then increases, reflecting the principle that it more likely to observe publications made by authors with a more extensive publication record.}

Both these aspects, i.e. non-monotonicity as well as the temporal heterogeneity of an effect, would not be possible to capture with a purely linear model. 
{Other approaches can be used to study the time-evolution of the impact of drivers on event rates. However, unlike the method proposed by \citet{kamalabad2023point}, we do not need to explicitly test for change points, as such variations are captured by the smooth function of time included in each effect. However, our approach assumes that the time function is smooth, so sudden changes in the effects -- permissible under the change point detection method -- are not easily accommodated in our approach. This is generally not an issue for publication mechanisms, particularly given the yearly granularity of the data. 
\citet{uzaheta2024modeling} considered the use of time-varying coefficients due to changes in network state in the context of DyNAMs. As they require  the assumption of piecewise constant inter-arrival times, the rate of state switching (causing a change in parameters) must be lower than the rate of relational events. This can be a limiting assumption, particularly in the presence of time-varying exogenous effects, especially if unmeasured.}

{Although we propose tensor product smoother, other approaches exist for fitting jointly time-varying non-linear effects, e.g., using an isotropic smooth function of time and the covariate. However, isotropic smooths can perform poorly when the covariates are on different scales \citep{wood2006low}. Although we transformed both time and the covariate to the interval \([0,1]\) to mitigate such problems, the employed tensor product smooth approach is generally safer, especially when applying the model directly to data without intermediate transformations.}

{If prior knowledge is available about the functional form of an effect, the estimation method can be adapted using shape constraints. SCAMs \citep{pya2015shape} extend GAMs by allowing component smooth functions to satisfy monotonicity or convexity constraints. This is achieved through a modified spline basis in which the constraints are enforced via a reparameterization of the spline coefficients.}

We showed how the dynamics related to specific drivers of the hyperevent data-generating process can be represented using two-dimensional heatmaps. What is relevant about this approach is that each heatmap reflects the most appropriate functional form according to the data. As for time-varying and non-linear effects, when dealing with a jointly time-varying non-linear effect, we let the contribution of each covariate take a data-driven shape. 
This increased flexibility comes at a computational cost. For each covariate, we need to estimate \(L \times Q\), where \(Q\) is the basis dimension related to non-linearity of the covariate and \(L\) is the basis dimension related to time. In the linear approach, for one covariate, we just need a single slope parameter. When dealing with many covariates, as in high-dimensional settings, this can pose a significant challenge -- especially when estimation relies on inverting the Hessian matrix. It would therefore be important to adapt stochastic gradient descent techniques, such as those proposed in \citet{filippi2024stochastic}, to relational hyperevent models and, in particular, to these types of flexible effect specifications.

{Although we employ relational hyperevent modelling to study the dynamics of publication events, the recorded publication date does not necessarily reflect when the underlying collaboration actually occurred, nor when authors became aware of relevant prior work. Substantial delays may arise for many reasons, including journal publication frequency, editorial and reviewer availability, and the number and duration of review rounds. Moreover, in more recent times researchers often learn about new work through online pre-prints well before formal publication dates. As a result, recorded publication dates may be delayed relative to the true timing of scientific activity. Because relational hyperevent models explain publication dynamics as a function of previous collaborations and state of the field, such timing discrepancies may have practical implications. In particular, delayed publication can make past events appear less temporally proximate than they actually were, weakening recency effects in the observed data. Delays may also reorder the hyperevents themselves. The yearly time granularity of the data -- although it may render several events as simultaneous when they are not -- helps mitigating some of the issues due to the delay.
}

    % CLOSING SECTIONS AND REFERENCES
    \section{Acknowledgments}

This work was supported by the Swiss National Science Foundation [grant 192549]; and the Deutsche Forschungsgemeinschaft (DFG) [Project No. 555455503].

\section{Declaration of interest}
None.

\section{Author contributions}
\textbf{MB}: Conceptualization, Methodology, Software, Formal analysis, Writing - Original Draft; 
\textbf{JL}: Conceptualization, Methodology, Software, Resources, Data Curation, Writing - Review \& Editing, Funding acquisition;
\textbf{EW}: Conceptualization, Methodology, Writing - Review \& Editing, Funding acquisition, Supervision; 

\newpage
\bibliographystyle{plainnat}  
\bibliography{_REFERENCES_.bib}

\end{document}